\documentclass[aps,twocolumn,float,prd,psfig,amsmath,amssymb]{revtex4}

\usepackage[dvipdfmx]{graphicx}
\usepackage{dcolumn}
\usepackage{bm}
\usepackage{braket}
\usepackage{breqn}
\usepackage{color}

\newcommand{\add}[1]{\textcolor{red}{#1}}

\begin{document}


\title{Correlation analysis for isotropic stochastic gravitational wave backgrounds with maximally allowed  polarization degrees}


\author{Hidetoshi Omiya}
\email{omiya@tap.scphys.kyoto-u.ac.jp}
\author{Naoki Seto}
\email{seto@tap.scphys.kyoto-u.ac.jp}
\affiliation{Department of Physics$,$ Kyoto University$,$ Kyoto 606-8502$,$ Japan}


\date{\today}

\begin{abstract}

 We study correlation analysis for  monopole components of stochastic gravitational wave  backgrounds,  including the maximally allowed polarization degrees. We  show that, for typical detector networks,  the correlation analysis can probe virtually five spectra: three for the intensities of the tensor, vector, and scalar modes and two for the chiral asymmetries of the tensor and vector modes.  The chiral asymmetric signal for the vector modes has been left untouched so far.   In this paper, we derive the overlap reduction function for this signal and thus complete the basic ingredients required for widely dealing with polarization degrees. We comprehensively analyze the geometrical properties of all the five overlap reduction functions.  In particular, we point out the importance of  reflection transformations for configuring preferable networks in the future. 
\end{abstract}


\maketitle

\section{Introduction}

A stochastic gravitational wave background is one of the important observational targets in the near future. It can be generated by cosmological processes \cite{Starobinsky:1979ty, PhysRevLett.99.221301, Kamionkowski:1993fg, Caprini:2007xq}, and  has a potential to probe unknown physics in the early universe (see \cite{Maggiore:1999vm,Romano:2016dpx,Christensen:2018iqi,Kuroyanagi:2018csn} for reviews). When searching for a cosmological background, our primary target would  be its  isotropic (monopole)  component.

The polarization states are basic characteristics of a background, and it would be interesting to discuss how well we can observationally extract related information.  
In fact, gravitational waves can take at most  six polarization patterns (see \cite{Will:1993ns} for the geometrical explanation of the patterns). In General Relativity (GR), we only have the two tensor (T) modes known as the $+$ and $\times$ patterns. However,  alternative theories of gravity allow the existence of the additional four modes: the two vector (V) modes (the $x$ and $y$ patterns) and the two scalar (S) modes (the $b$ and $l$ patterns).  By detecting these vector and scalar modes of  a background, we can probe a violation of GR (see for example \cite{Nishizawa:2009bf, Nishizawa:2009jh, Cornish:2017oic, LIGOScientific:2019vic}). For the tensor and vector sectors, we could introduce the circular polarization bases, instead of the familiar linear ones.  The asymmetry between the right- and left-handed patterns of a background would be a strong evidence for a parity violation process \cite{Lue:1998mq,Seto:2006hf,Kato:2015bye,Smith:2016jqs,Domcke:2019zls,Belgacem:2020nda} (see also \cite{Alexander:2004us, Satoh:2007gn,Adshead:2012kp,Kahniashvili:2005qi,Ellis:2020uid} for generations). 

The correlation analysis is a  powerful approach for detecting weak stochastic background signals under the presence of detector noises \cite{Christensen:1992wi, Flanagan:1993ix, Allen:1997ad}.  When considering only the monopole components, under the low frequency approximation, we can show that  the correlation analysis can virtually probe the five spectra ($I_T,I_V,I_S,W_T,W_V$) characterizing  the polarization states. The three spectra $(I_T,I_V,I_S)$ represent  the total intensities of the tensor, vector, and scalar modes, respectively. The other two spectra $(W_T,W_V)$ show the chiral asymmetries of the tensor and vector modes which are usually referred to as the Stokes \lq\lq{}$V$\rq\rq{} parameter. 
In this paper, we used the notation $W$ instead of the conventional one  \lq\lq{}$V$\rq\rq{}, not to confuse with the abbreviation $V$ which  exclusively represents the vector  modes. 

Here, we should notice that the chiral spectra $(W_T,W_V)$ change their signs with respect to the parity transformation. On the other hand, the total intensities  ($I_T,I_V,I_S$) are invariant under the  parity transformation. Therefore, we contrastively call the formers by the parity odd spectra and later by the parity even spectra.

The overlap reduction functions (ORFs) describe the correlated responses of the two detectors to the five spectra. We denote them by $\gamma^{I_P} (P=T,V,S)$ and $\gamma^{W_P} (P=T,V)$.  These ORFs  play  key  roles in the correlation analysis.  The analytic expressions for  four functions  ($\gamma^{I_T},\gamma^{I_V},\gamma^{I_S}, \gamma^{W_T}$) can be found in the literatures \cite{Flanagan:1993ix, Allen:1997ad, Nishizawa:2009bf, Nishizawa:2009jh, Seto:2008sr}. However,  the parity asymmetric vector spectrum $W_ V$ and the associated ORF $\gamma^{W_V}$ have been left untouched so far. 

In this paper, we  derive the analytic expression for the remaining function $\gamma^{W_V}$ and make comparative discussions on the five ORFs, paying special attention to behaviors under the parity and reflection transformations. We will see that these transformations are crucial for optimizing the sensitivity to the asymmetric components $(W_T,W_V)$ and their isolation from the symmetric  components ($I_T,I_V,I_S$). These results would be useful for designing future detector networks, including space interferometers.

This paper is organized as follows. In Sec. \ref{sec:2}, we review the possible polarization states of a stationary and isotropic  gravitational wave background. We will argue that the monopole pattern of a background is effectively characterized by the five spectra $(I_T,I_V,I_S,W_T,W_V)$. In Sec. \ref{sec:3}, we discuss the correlation analysis of a stochastic gravitational wave background and newly derive the analytic expression for  $\gamma^{W_V}$. In Sec. \ref{sec:4}, we focus on the networks which are insensitive to the parity even spectra ($I_T,I_V,I_S$) for solely detecting the odd ones $(W_T,W_V)$. In Sec. \ref{sec:5}, we summarize our results and discuss possible applications of our study.  

\section{Polarization of a stochastic gravitational wave background}\label{sec:2}

First, we describe the polarization decomposition of a stochastic gravitational wave background, specifically focusing on the vector modes. Since our universe is highly isotropic and homogeneous, we set an isotropic background as our primary target. In addition, considering that the observed propagation speed $v_g$ of gravitational waves  is nearly the same as the speed of light $c$ \cite{TheLIGOScientific:2017qsa}, we hereafter assume $v_g = c$. 

We formally apply the plane wave expansion for the metric perturbation $h_{ij}$ induced by a gravitational wave background as
\begin{align}\label{eq:1}
\begin{aligned}
	h_{ij}(t,\bm{x}) = &\sum_{P} \int df \int d\bm{\Omega}\\
	& \times \tilde{h}_P(f,\bm{\Omega}) \bm{e}_{P,ij}(\bm{\Omega}) e^{-2\pi i f (t - \bm{\Omega} \cdot \bm{x}/c)}~.
\end{aligned}
\end{align}
Here, $\bm{\Omega}$ is the unit vector on the two sphere parametrized by
\begin{align}
	\bm{\Omega} &= \left(
	\begin{array}{c}
	\sin\theta \cos\phi\\
	\sin\theta\sin\phi\\
	\cos\theta
	\end{array}
	\right)
\end{align}
in the cartesian coordinate. Note that we normalized the integral measure $d\bm{\Omega}$ by  $\int d\bm{\Omega} = 4 \pi$.  

In Eq. (\ref{eq:1}), the expression $e_{ij,P}(\Omega)$ represents the polarization tensor and $\tilde{h}_P$ is the mode coefficient. The subscript $P$ denotes the polarization modes of a gravitational wave and takes the following six patterns  $P = +,\times, x, y, b,$and  $l$  in the most general case \cite{Will:1993ns}. The $+$- and $\times$-patterns are the usual tensor (T) modes predicted by GR. On the other hand, the remaining modes do not appear in GR. The $x$- and $y$-patterns  are called the vector (V) modes and the $b$- and $l$-patterns are the  scalar (S) modes. The corresponding polarization tensors are given by
\begin{align}\label{eq:3}
\begin{aligned}
	\bm{e}_{+} &= \bm{m} \otimes \bm{m} - \bm{n}\otimes \bm{n}~, & \bm{e}_{\times} &= \bm{m} \otimes \bm{n} + \bm{n}\otimes \bm{m}~,\\
	 \bm{e}_{x} &= \bm{\Omega} \otimes \bm{m} + \bm{m}\otimes \bm{\Omega}~,& \bm{e}_y &= \bm{\Omega} \otimes \bm{n} + \bm{n}\otimes \bm{\Omega}~,\\
	 \bm{e}_b &= \sqrt{3}(\bm{m} \otimes \bm{m} + \bm{n}\otimes \bm{n})~,& \bm{e}_l &= \sqrt{3}(\bm{\Omega}\otimes\bm{\Omega})~,
\end{aligned}
\end{align}
where the unit transverse vectors $\bm{m}$ and $\bm{n}$ are given by
\begin{align}
	\bm{m} &= \left(
	\begin{array}{c}
	\cos\theta \cos\phi\\
	\cos\theta\sin\phi\\
	-\sin\theta
	\end{array}
	\right)~, &
	\bm{n} &= \left(
	\begin{array}{c}
	-\sin\phi\\
	\cos\phi\\
	0
	\end{array}
	\right)~.
\end{align}
Here, the unconventional factor $\sqrt{3}$ of  the scalar modes are chosen for normalizing   the strain fluctuations  (see appendix A of \cite{Omiya:2020fvw}).

In Eq. \eqref{eq:1}, the mode coefficients $\tilde{h}_P$ are random quantities and their statistical properties are characterized by the power spectrum densities.  In  concrete terms, we first discuss the vector modes. Because the vector modes (the $x$- and $y$- patterns) can be regarded  as the massless spin-1 particles \cite{Will:1993ns}, we can characterize their polarization properties  similarly to the electromagnetic waves \cite{1979rpa..book.....R}. Therefore, the $2\times 2$ matrix (for $P,P' = x,y$) for their power spectra  is given by 
\begin{align}\label{eq:5}
\begin{aligned}
	\braket{\tilde{h}_{P}(f,\bm{\Omega})\tilde{h}_{P'}^*(f',\bm{\Omega'})}& = \frac{1}{2}\delta_{\Omega\Omega'}\delta(f-f')\\
	&\times \left(
	\begin{array}{cc}
		I_V+Q_V & U_V-iW_V\\
		U_V+iW_V & I_V-Q_V
	\end{array}
	\right)~,
\end{aligned}
\end{align}
with the ensemble average $\braket{\cdots}$ (omitting the $(f,\bm{\Omega})$ dependence existing in the right-hand-side). Here, $I_V,W_V, Q_V$, and $ U_V$ are the Stokes parameters \cite{1979rpa..book.....R}.  As already mentioned,  we used $W$ instead of the conventional choice $V$ (representing \lq\lq{}vector\rq\rq{} in this paper).

The physical meaning of $I_V$ and $W_V$  can be seen more clearly in the circular polarization bases $(\bm{e}^V_R,\bm{e}^V_L)$ rather than the linear polarization bases $(\bm{e}_x,\bm{e}_y)$. We can relate them  by
\begin{align}
	\bm{e}^V_R &= \frac{1}{\sqrt{2}}\left(\bm{e}_x + i \bm{e}_y\right)~, & \bm{e}^V_L &= \frac{1}{\sqrt{2}}\left(\bm{e}_x - i \bm{e}_y\right)~.
\end{align}
Using these relations, the corresponding mode coefficients in the circular polarization bases are given by
\begin{align}\label{eq:7}
	\tilde{h}^{V}_L (f,\bm{\Omega})&= \frac{1}{\sqrt{2}}\left(\tilde{h}_x(f,\bm{\Omega}) + i \tilde{h}_y(f,\bm{\Omega})\right)~, \\
	\label{eq:8}
	 \tilde{h}^{V}_R(f,\bm{\Omega}) &= \frac{1}{\sqrt{2}}\left(\tilde{h}_x(f,\bm{\Omega}) - i \tilde{h}_y(f,\bm{\Omega})\right)~.
\end{align}
Then, from Eqs. \eqref{eq:5}, \eqref{eq:7}, and \eqref{eq:8} we obtain
\begin{align}\label{eq:9}
\begin{aligned}
	I_V &= \braket{\tilde{h}^V_R\tilde{h}_R^{V*}} + \braket{\tilde{h}^V_L \tilde{h}_L^{V*}}~, \\
	 W_V &= \braket{\tilde{h}^V_R \tilde{h}_R^{V*}} - \braket{\tilde{h}^V_L \tilde{h}_L^{v*}}~,\\
	Q_V &= \braket{\tilde{h}^V_R \tilde{h}_L^{V*}} + \braket{\tilde{h}_L \tilde{h}_R^{V*}}~, \\
	U_V &= i(\braket{\tilde{h}_R^V \tilde{h}_L^{V*}} - \braket{\tilde{h}_L^V \tilde{h}_R^{V*}})~,
\end{aligned}
\end{align}
by omitting the delta functions and the $(f,\bm{\Omega})$ dependence. From these expressions, we see that $I_V$ and $W_V$ respectively characterize the total and asymmetry  of the amplitudes of  the right and the left handed waves. We can also confirm that  the combinations $Q_V \pm i U_V$ characterize the linear polarization \cite{1979rpa..book.....R}.

As commented earlier, the polarization patterns of the vector modes are essentially the  same as the electromagnetic waves. Therefore, if we rotate the transverse  coordinate around the propagation direction $\bm{\Omega}$  by the angle $\alpha$, the left- and right- polarization modes transform as 
\begin{align}
	\tilde{h}^V_L &\to e^{-i\alpha} \tilde{h}^V_L~,\\
	\tilde{h}^V_R &\to e^{+i\alpha} \tilde{h}^V_R~.
\end{align}
From Eq. \eqref{eq:9}, we correspondingly obtain
\begin{gather}
\begin{aligned}
	&I_V \to I_V~,  W_V \to W_V~, \\
	&Q_V \pm i U_V \to e^{\pm 2i\alpha}(Q_V \pm i U_V)~.
\end{aligned}\label{eq:12}
\end{gather}
We observe that  $I_V$ and $W_V$ are spin-0 and $Q_V \pm i U_V$ are  spin-2. Below, we focus our study on the isotropic and stationary  background with no preferred spatial direction and orientation. Then, for the correlation between the mode coefficients, we only need to keep the monopole components of the spin-0 combinations. This is because a higher spin combination ({\it e.g.} $Q_V \pm i U_V$) introduces a specific orientation and should vanish for an  isotropic background. Accordingly, we hereafter  put $  I_V(f,\bm{\Omega})= I_V(f)$ and $W_V(f,\bm{\Omega})=W_V(f)$, ignoring the angular pattern. 

For the tensor modes, the power spectra are analogous to the vector modes. Indeed, the $2\times 2$ matrix (for $P,P' = +,\times$) for the tensor power spectra is given by
\begin{align}\label{eq:13}
\begin{aligned}
	\braket{\tilde{h}_{P}(f,\bm{\Omega})\tilde{h}_{P'}^*(f',\bm{\Omega'})}& = \frac{1}{2}\delta_{\Omega\Omega'}\delta(f-f')\\
	&\times \left(
	\begin{array}{cc}
		I_T+Q_T & U_T-iW_T\\
		U_T+iW_T & I_T-Q_T
	\end{array}
	\right)~
\end{aligned}
\end{align}
\cite{Seto:2008sr}. But here, we should recall that  the tensor modes are spin-2. Then, the Stokes parameters are transformed  similarly to the vector modes, as shown in Eq. \eqref{eq:12}  with the factor $e^{\pm4 i \alpha}$ for the linear polarization ($Q_T \pm i U_T$). Therefore, we keep only $ I_T(f,\bm{\Omega})=I_T(f)$ and $W_T(f,\bm{\Omega})=W_T(f)$ for an isotropic background, as in the case of the vector modes.

The mode coefficients for the scalar modes are transformed as spin-0 particles. We put their covariance matrix ($P,P' = b,l$) as
\begin{align}\label{eq:14}
\begin{aligned}
	\braket{\tilde{h}_{P}(f,\bm{\Omega})\tilde{h}_{P'}^*(f',\bm{\Omega'})} =& \frac{1}{2}\delta_{\Omega\Omega'}\delta(f-f')\\
	&\times \left(
	\begin{array}{cc}
		I_b & C_S\\
		C_S^* & I_l
	\end{array}
	\right)
\end{aligned}
\end{align}
taking into account the possible off-diagonal (correlation) terms. However, as long as the low frequency approximation is valid, the two modes are observationally degenerated (as shown in the next section). As a result, the correlation analysis can  probe only the mean amplitude defined by
\begin{align}\label{eq:15}
 I_S \equiv  \frac{1}{2}\left(I_b + I_l - C_S - C_S^*\right)~.
\end{align}
The mechanism behind this expression will be explained also in the next section.

One might be interested in  the correlation between different polarization modes, such as the $T-V$, $T-S$, and $V-S$ pairs. However, they cannot produce  spin-0 combinations and should vanish for an isotropic background.

To summarize, there are at most five spectra ($I_T,I_V,I_S,W_T$ and $W_V$) that effectively characterize a stationary and isotropic gravitational wave background (under the low frequency approximation). The spectra $I_T,I_V,$ and $I_S$  represent the total intensity of the tensor, vector, and scalar modes respectively.  In contrast, the spectra $W_T$ and $W_V$ are the odd parity component and probe the parity violation process for  the tensor and vector polarizations.

\section{Overlap reduction functions}\label{sec:3}

The correlation analysis is a powerful approach for detecting a gravitational wave background \cite{Christensen:1992wi, Flanagan:1993ix, Allen:1997ad}. The ORFs are its key elements and characterize the correlated response of two detectors to an isotropic background.  As mentioned earlier, we have, in total,  the  five monopole spectra. The ORF for $I_T$ was first discussed in \cite{Christensen:1992wi, Flanagan:1993ix},  for $W_T$  in \cite{Seto:2006dz,Seto:2007tn}, and for $I_V$ and $I_S$ in \cite{Nishizawa:2009bf, Nishizawa:2009jh}.   However, the function for the remaining one $W_V$ had not been studied so far.

In this section, we discuss the ORFs, paying special attention to the unexplored one $W_V$ in relation to the analog one $W_T$. For systematically handling intermediate calculations, we will also introduce the new orthogonal tensorial bases, utilizing the underlying geometrical symmetry. 

\subsection{General formulation}\label{sec:3A}

Let us consider the interaction of a detector $A$ with a gravitational wave background. Under the low frequency approximation valid for  $f \ll (2\pi L/c)^{-1}$  with the detector arm length $L$,  the response of the detector can be modeled as \cite{Forward:1978zm}
\begin{align}\label{eq:16}
	h_A(f) = D_{A}^{ij} \tilde{h}_{ij}(f, \bm{x}_A)~.
\end{align}
Here, $\tilde{h}_{ij}(f,\bm{x}_A)$ is the metric perturbation of the background at the detector. We also defined   the detector tensor  
$\bm{D}_A$ by
\begin{align}\label{eq:17}
	\bm{D}_{A} = \frac{\bm{u}_A \otimes \bm{u}_A - \bm{v}_A \otimes \bm{v}_A}{2}
\end{align}
where $\bm{u}_A$ and $\bm{v}_A$ are the unit vectors representing the two  arm directions.

By cross-correlating two noise independent detectors, one can distinguish a stochastic background from random detector noises. We define the expectation value of the correlated signal for two detectors $A$ and $B$ as 
\begin{align}\label{eq:18}
	C_{AB}(f) \equiv \braket{h_A(f) h_B^*(f)}~,
\end{align}
again omitting the apparent delta function. Using Eqs. \eqref{eq:1}, \eqref{eq:5}, \eqref{eq:13}, \eqref{eq:14}, and \eqref{eq:16}, and leaving only the monopole components, we obtain
\begin{align}\label{eq:19}
\begin{aligned}
	C_{AB}(f) = \frac{4\pi}{5}& D_{A}^{ij} D_{B}^{kl}\left(\sum_{P=T,V,S}\Gamma^{I_P}_{ijkl}(f) I_P(f) \right.\\
	&\left. + \sum_{P=T,V}\Gamma^{W_P}_{ijkl}(f) W_P(f)\right)~,
\end{aligned}
\end{align}
with 
\begin{align}\label{eq:20}
	\Gamma^{I_T}_{ijkl} &\equiv \frac{5}{8 \pi} \int d\bm{\Omega} (e_{+,ij} e_{+,kl} + e_{\times,ij} e_{\times,kl}) e^{i y \bm{\Omega}\cdot \hat{\bm{d}}}~,\\
	\label{eq:21}
	\Gamma^{I_V}_{ijkl} &\equiv \frac{5}{8 \pi} \int d\bm{\Omega} (e_{x,ij} e_{x,kl} + e_{y,ij} e_{y,kl}) e^{i y \bm{\Omega}\cdot \hat{\bm{d}}}~,\\
	\label{eq:22}
	\Gamma^{I_S}_{ijkl} &\equiv \frac{5}{8 \pi} \int d\bm{\Omega} (e_{b,ij} e_{b,kl} + e_{l,ij} e_{l,kl}) e^{i y \bm{\Omega}\cdot \hat{\bm{d}}}~,\\
	\label{eq:23}
	\Gamma^{W_T}_{ijkl} &\equiv -\frac{5 i}{8 \pi} \int d\bm{\Omega} (e_{+,ij} e_{\times,kl} - e_{\times,ij} e_{+,kl}) e^{i y \bm{\Omega}\cdot \hat{\bm{d}}}~,\\
	\label{eq:24}
	\Gamma^{W_V}_{ijkl} &\equiv -\frac{5 i}{8 \pi} \int d\bm{\Omega} (e_{x,ij} e_{y,kl} - e_{y,ij} e_{x,kl}) e^{i y \bm{\Omega}\cdot \hat{\bm{d}}}~.
\end{align}
In these expressions, we introduced the unit vector $\hat{\bm{d}} = (\bm{x}_A - \bm{x}_B)/d$ with $d \equiv |\bm{x}_A - \bm{x}_B|$  and $y =2\pi f d/c$. 
The tensors $\Gamma_{ijkl}^{I_P,W_P}$ should be completely determined by $y$ and $\hat{d}$. 

Here we comment on the degeneracy between the two scalar patters ($b$ and $l$). Note that, the sum of the polarization tensor for these two patterns is
\begin{align}
	e_{b,ij} + e_{l,ij} = \sqrt{3} \delta_{ij}~.
\end{align}
Since the detector tensor $D_{ij}$ is traceless, we identically have
\begin{align}\label{eq:add1}
	e_{b,ij}D_{ij} = - e_{l,ij}D_{ij}~.
\end{align}
With the identity Eq. \eqref{eq:add1},   the cross correlation between the scalar modes can be evaluated as 
\begin{widetext}
\begin{align}
	\braket{h_A(f)h_B^*(f)}|_{\rm scalar} &= D_{A,ij}D_{B,kl} \int d\bm{\Omega}\left(e_{b,ij}e_{b,kl}\frac{I_b}{2} + e_{l,ij}e_{l,kl}\frac{I_l}{2} + e_{b,ij}e_{l,kl}\frac{C_S}{2} + e_{l,ij}e_{b,kl}\frac{C_S^*}{2}\right) e^{i y \bm{\Omega} \cdot \hat{\bm{d}}}\\
		\label{eq:A4}
		&=  \frac{1}{2} D_{A,ij}D_{B,kl}\left(\int d\bm{\Omega} \left(e_{b,ij}e_{b,kl} + e_{l,ij}e_{l,kl} \right) e^{i y \bm{\Omega} \cdot \hat{\bm{d}}}\right) \frac{I_b + I_l - C_S -C_s^*}{2}~.
\end{align}
\end{widetext}
As a result, under the low frequency approximation, only the combination $I_S$ in Eq. \eqref{eq:15} appears for the cross correlation.

By contracting the tensors $\Gamma_{ijkl}^{I_P}$ and  $\Gamma_{ijkl}^{W_P}$  with detector tensors $D_A^{ij}$ and  $D_B^{ij}$, we obtain the formal expression of the ORFs as
\begin{align}\label{eq:27}
	\gamma^{I_P}_{AB}&(f) \equiv D_{A}^{ij}D_B^{kl} \Gamma^{I_P}_{ijkl}~,\\
	\label{eq:28}
	\gamma^{W_P}_{AB}&(f) \equiv D_{A}^{ij}D_{B}^{kl} \Gamma^{W_P}_{ijkl}~.
\end{align}
Then the cross correlation \eqref{eq:19} can be written as
\begin{align} \label{c18}
\begin{aligned}
	C_{AB}(f) = \frac{4\pi}{5}&\left(\sum_{P=T,V,S}\gamma^{I_P}_{AB}(f) I_P(f) \right.\\
	&\left. + \sum_{P=T,V}\gamma^{W_P}_{AB}(f) W_P(f)\right)~.
\end{aligned}
\end{align}
The functions $\gamma^{I_P}_{AB}$ and $\gamma^{W_P}_{AB}$ clearly characterize the correlated response of the detectors to the corresponding background spectra. Following the classification of the spectra, we call $\gamma^{I_P}_{AB} (P=T,V,S)$ by the parity even ORFs and $\gamma^{W_P}_{AB}(P=T,V)$ by the parity odd ones.

From Eqs. \eqref{eq:20} - \eqref{eq:24}, we can identify the symmetric properties of  the tensors $\Gamma_{ijkl}^{I_P,W_P}$. From the definition of the polarization bases (see Eq. \eqref{eq:3}), the tensor $\Gamma_{ijkl}^{I_P,W_P}$ are symmetric under exchange of indices,
\begin{align}\label{eq:symm1}
	\Gamma^{I_P,W_P}_{ijkl} = \Gamma^{I_P,W_P}_{jikl} = \Gamma^{I_P,W_P}_{ijlk}
\end{align}
Also, it is easy to confirm that the parity even ones (\eqref{eq:20} - \eqref{eq:22}) satisfy
\begin{align}\label{eq:29}
	\Gamma^{I_P}_{ijkl} &= \Gamma^{I_P}_{klij}~,
\end{align}
while we have
\begin{align}\label{eq:30}
	\Gamma^{W_P}_{ijkl} &= -\Gamma^{W_P}_{klij}~
\end{align}
for the odd ones (\eqref{eq:23} and \eqref{eq:24}). In the next subsection (and appendix A),  we extensively use these properties for deriving analytic expressions of the ORFs.

\subsection{Symmetries to Transformations}\label{3q}

At this point, we briefly discuss responses of various quantities to the three-dimensional parity transformation. As commented earlier, it interchanges the right- and left-handed waves, keeping the scalar modes invariant.  We have the correspondences for the spectra
\begin{align}
I_P\rq{}=I_P,~~W_P\rq{}=-W_P \label{p1}
\end{align}
with the prime attached to the transformed quantities.
Since the correlation product is unchanged with the parity transformation, we also have
\begin{align}
C_{AB}\rq{}(f)=C_{AB}(f) , \label{p2}
\end{align}
and resultantly obtain
\begin{align}
{\gamma_{AB}^{I_P}}\rq{} (f)={\gamma_{AB}^{I_P}}(f),~~{\gamma_{AB}^{W_P}}\rq{} (f)=-{\gamma_{AB}^{W_P}}(f).\label{p3}
\end{align}
Meanwhile, we can easily confirm 
\begin{align}
\bm{D}_{A}\rq{}=\bm{D}_{A} ,~~\bm{D}_{B}\rq{}=\bm{D}_{B} \label{p4}.
\end{align}
Then, we have 
\begin{align}
{\Gamma^{I_P}_{ijkl}}\rq{} =\Gamma^{I_P}_{ijkl},~~{\Gamma^{W_P}_{ijkl}}\rq{} =-\Gamma^{W_P}_{ijkl}.\label{p5}
\end{align}

Next, we discuss a mirror transformation (reflection) with respect to a single plane. In fact, the parity transformation is generated by the consecutive operations of a mirror transformation and an associated spatial rotation (e.g. reflection at the $yz$-plane and  rotation around the $x$-axis by the angle $180^\circ$). Since we are dealing with an isotropic background, a spatial rotation plays no role in correlation analysis. Therefore, for a mirror transformation at an arbitrary plane, we have the correspondences identical to  (\ref{p1})-(\ref{p3}) (but not generally (\ref{p4}) and (\ref{p5})). 

\subsection{Construction of Parity Odd ORFs}\label{sec:3B}

Now we derive  analytic expressions for the parity odd ORFs $\gamma^{W_T}_{AB}$ and $\gamma^{W_V}_{AB}$. Our basic strategy is to apply the irreducible decomposition on the rank-4 tensors $\Gamma^{W_P}_{ijkl}$. This is a systematic extension of the approach applied in Flanagan \cite{Flanagan:1993ix}, but has not been used in the literature. In appendix \ref{sec:AppA}, we discuss the parity even ORFs,  following a similar procedure.

First, we specify the possible tensors which  can be used for composing $\gamma^{W_T}_{AB}$. As already mentioned, the unit vector $\hat{\bm{d}}$ should be the unique candidate for the vector.  In addition, since the polarization tensors transform under the special orthogonal group ${\rm SO}(3)$, the tensors $\Gamma^{W_P}_{ijkl}$ should also transform under ${\rm SO}(3)$. Correspondingly, in addition to $\hat{\bm{d}}$, we may use $\delta_{ij}$ and $\epsilon_{ijk}$. Therefore, the basic building blocks of the $\Gamma_{ijkl}$ must be the  three tensors below
\begin{align}\label{eq:31}
&\delta_{ij}~, & M^{0}_{ij} &\equiv \hat{d}_i \hat{d}_j - \delta_{ij}/3~,& \omega_{ij} &\equiv \epsilon_{ijk}\hat{d}_k~.
\end{align}
Note that $M^{0}$ is traceless and symmetric, while $\omega$ is anti-symmetric.

Following the general procedure for the irreducible decomposition of ${\rm SO}(N)$ tensor \cite{hamermesh1989group}, we construct the tensorial bases for the rank-4 tensors satisfying Eqs. \eqref{eq:symm1} and \eqref{eq:30}. The relevant tensors should be the following ones:
\begin{align}\label{eq:37}
&\tilde{H}^{0}_{ijkl} = \frac{1}{2}\left(\delta_{ij} M^0_{kl} - M^0_{ij} \delta_{kl}\right)~,\\
	\label{eq:38}
		&\tilde{K}^{0}_{ijkl} = \frac{1}{2\sqrt{10}}\left(\delta_{ik}\omega_{jl} + \delta_{il}\omega_{jk}+ \delta_{jk}\omega_{il} + \delta_{jl}\omega_{ik}\right)~,\\
	\label{eq:39}
		&\begin{aligned}
	\tilde{F}^{0}_{ijkl} = & \frac{1}{4}\sqrt{\frac{5}{2}}\left(\omega_{ik}M^{0}_{jl} + \omega_{il}M^{0}_{jk}+ \omega_{jk}M^{0}_{il} + \omega_{jl}M^{0}_{ik}\right)\\
	& + \frac{1}{3}\tilde{K}^{0}_{ijkl}~,
	\end{aligned}
\end{align}
These  tensors are orthonormal in the following sense
\begin{gather}\label{eq:340}
	\tilde{F}^0_{ijkl} \tilde{H}^0_{ijkl} = \tilde{F}^0_{ijkl} \tilde{K}^0_{ijkl} = \tilde{H}^0_{ijkl} \tilde{K}^0_{ijkl} =0~,\\
	\label{eq:41}
	\tilde{F}^0_{ijkl} \tilde{F}^0_{ijkl} = \tilde{H}^0_{ijkl} \tilde{H}^0_{ijkl} = \tilde{K}^0_{ijkl} \tilde{K}^0_{ijkl} =1~.
\end{gather}
Note that $\tilde{F}^0$ is traceless 
\begin{align}
	\tilde{F}^0_{iijk} = \tilde{F}^0_{ijik} = \dots = \tilde{F}^0_{jkii} = 0~,
\end{align}
which is required by the irreducible decomposition.

Using the tensors \eqref{eq:37},\eqref{eq:38}, and \eqref{eq:39}, we can expand $\Gamma^{W_P}$ as 
\begin{align}
	\Gamma^{W_P}_{ijkl} &= \tilde{\rho}^P_{\tilde{F}} \tilde{F}^{0}_{ijkl}  + \tilde{\rho}^P_{\tilde{K}} \tilde{K}^0_{ijkl} + \tilde{\rho}^P_{\tilde{H}} \tilde{H}^{0}_{ijkl}~.
\end{align}
The orthonormal nature of the basis allows us to obtain the expansion coefficients by simply contracting the tensors with $\Gamma^{W_P}_{ijkl}$:
\begin{align}
	\tilde{\rho}^P_{\tilde{F}} &= \tilde{F}^{0}_{ijkl}\Gamma^{W_P}_{ijkl}~,\\
	\tilde{\rho}^P_{\tilde{H}} &= \tilde{H}^{0}_{ijkl}\Gamma^{W_P}_{ijkl}~,\\
	\tilde{\rho}^P_{\tilde{K}} &= \tilde{K}^{0}_{ijkl}\Gamma^{W_P}_{ijkl}~.
\end{align}
After some elementary calculations, we obtain the coefficients
\begin{align}
	(\tilde{\rho}^T_{\tilde{F}},\tilde{\rho}^T_{\tilde{K}},\tilde{\rho}^T_{\tilde{H}}) &= \sqrt{10}(j_3(y),2 j_1(y),0)~,\\
	(\tilde{\rho}^V_{\tilde{F}},\tilde{\rho}^V_{\tilde{K}},\tilde{\rho}^V_{\tilde{H}}) &= \sqrt{10}(-2 j_3(y),j_1(y),0)~,
\end{align}
where $j_n(y)$ are the spherical Bessel functions. Notice that we identically have $\rho^P_{\tilde{H}}=0$.
This straightforwardly  follows from the oddness of the tensors $\Gamma^{W_P}_{ijkl}$ with respect to parity transformation.  We must  have an odd power of  $\hat{\bm{d}}$, in contrast to  $\tilde{H}^{0}_{ijkl}$.

Using the expressions presented so far, we have the analytic expressions of $\gamma^{W_T}$ and $\gamma^{W_V}$ as
\begin{align}\label{eq:49}
	\gamma^{W_T}_{AB} &= \sqrt{10}\left(D^{\tilde{F}}_{AB} j_3(y) + 2 D^{\tilde{K}}_{AB} j_1(y)\right)~,\\
	\label{eq:50}
	\gamma^{W_V}_{AB} &= \sqrt{10}\left(-2 D^{\tilde{F}}_{AB} j_3(y) + D^{\tilde{K}}_{AB} j_1(y)\right)~.
\end{align}
Here, we defined
\begin{align}\label{eq:51}
	D^{\tilde{F}}_{AB} &\equiv D_{A,ij}D_{B,kl} \tilde{F}^0_{ijkl}~,\\
	\label{eq:52}
	D^{\tilde{K}}_{AB} &\equiv D_{A,ij}D_{B,kl} \tilde{K}^0_{ijkl}~.
\end{align}
We present the ready-to-use expressions for networks composed by two ground-based detectors in appendix \ref{sec:AppC}. In Fig. \ref{fig:1}, as typical examples,   we show the parity odd ORFs for the VIRGO-LIGO Hanford network.

\begin{figure}
\centering
\includegraphics[keepaspectratio, scale=0.55]{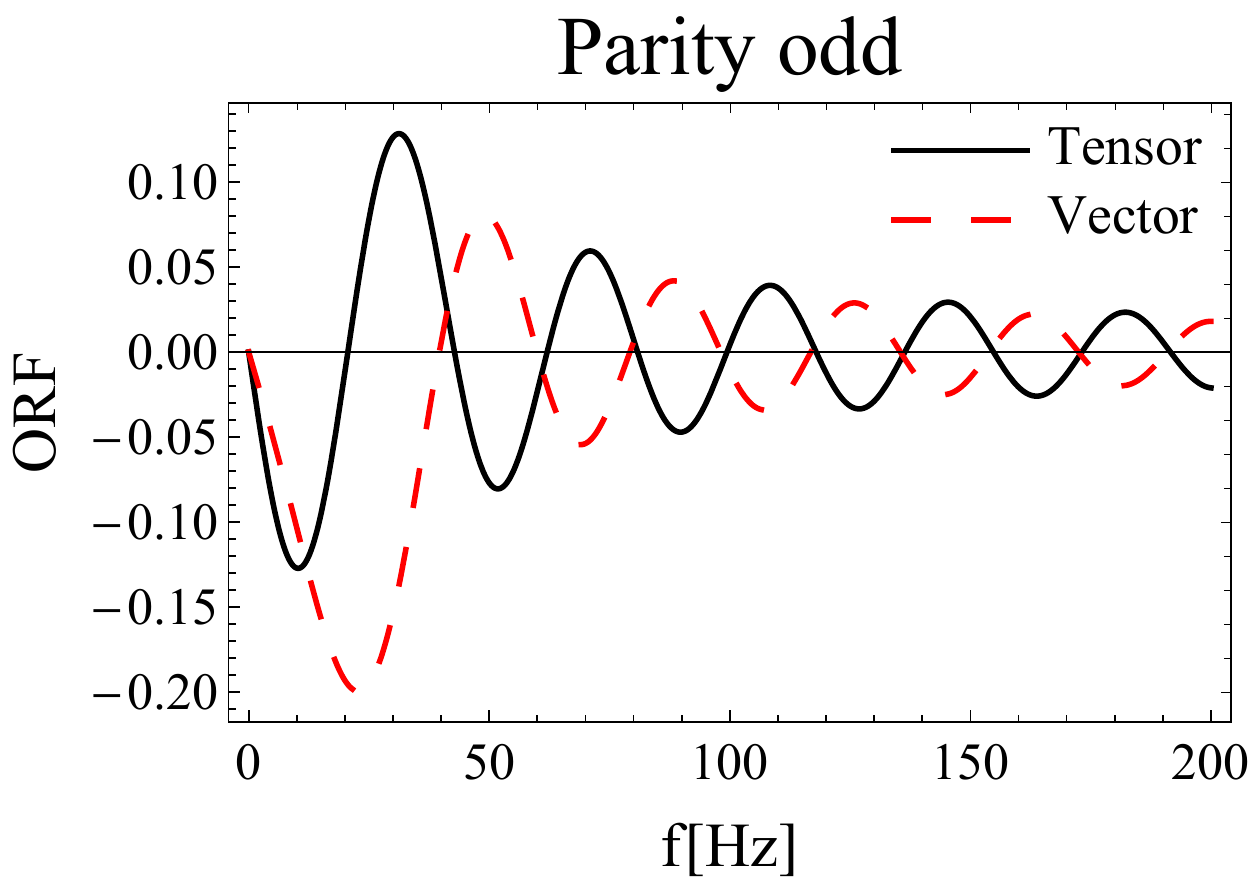}
\caption{The parity odd ORFs for the VIRGO-LIGO Hanford network. The black solid and red dashed lines correspond to  the tensor and  vector modes, respectively. The distance between two detectors is $d \sim 8.2\times 10^3 {\rm km}$.}
\label{fig:1}
\end{figure}

The asymptotic behaviors of the ORFs can be easily understood from Eqs. \eqref{eq:49} and \eqref{eq:50}. For the large frequency regime ($y = 2\pi f d/c \gg 1$), the spherical Bessel functions behave as
\begin{align}
	j_1(y) \underset{y\to \infty}{\to} - \frac{\cos y}{y}~,\\
	j_3(y) \underset{y\to \infty}{\to} \frac{\cos y}{y}~.
\end{align}
Then we have
\begin{align}
	\gamma^{W_T}_{AB} \underset{y\to \infty}{\to} \sqrt{10}\left(D^{\tilde{F}}_{AB} - 2 D^{\tilde{K}}_{AB}\right) \frac{\cos y}{y}~,\\
	\gamma^{W_V}_{AB} \underset{y\to \infty}{\to}\sqrt{10}\left(-2 D^{\tilde{F}}_{AB} - D^{\tilde{K}}_{AB}\right) \frac{\cos y}{y}~.
\end{align}
Thus, they oscillate with the frequency interval $c/d$ and the envelope $\propto 1/f$, as in Fig. \ref{fig:1}.

In the small frequency regime ($y \ll 1$), the spherical Bessel functions can be expanded as
\begin{align}
	j_1(y) \underset{y\to 0}{\to} \frac{y}{3}~,\\
	j_3(y) \underset{y\to 0}{\to} \frac{y^3}{105}~.
\end{align}
We then  have 
\begin{align}\label{eq:54}
	\gamma^{W_T}_{AB} \underset{y\to 0}{\to} \frac{2\sqrt{10}}{3}D^{\tilde{K}}_{AB}\, y~,\\
	\label{eq:55}
	\gamma^{W_V}_{AB} \underset{y\to 0}{\to}\frac{\sqrt{10}}{3}D^{\tilde{K}}_{AB}\, y~.
\end{align}
Therefore, the ORFs approach to zero in the small frequency regime as shown in Fig. \ref{fig:1}. 

These asymptotic behaviors indicate the blindness of the coincident detectors ($d = 0$ or equivalently  $y=0$) to the parity odd components.  We can understand this from the oddness of the function  $\gamma^{W_P}_{AB}$ with respect to the parity transformation (essentially the same as the discussion on $\rho^P_{\tilde{H}} = 0$ above).  In the next section, we discuss the responses  of the parity odd ORFs to a mirror transformation (reflection).

\section{Asymmetric  networks}\label{sec:4}

As shown in Eq. \eqref{eq:27}, the expectation value of the correlation product is a linear combination of the even and odd parity spectra.  Since the latter are closely related to the parity violation process, we would like to detect them without contamination by the even spectra.  In this section, using a mirror transformation, we discuss how to realize the desired  network with $\gamma_{AB}^{I_p}=0$ ($P=T,V$ and $S$). In the following, to simplify our expressions, we omit the subscript $AB$ (labels for detectors) and put $\gamma^{I_P}_{AB} = \gamma^{I_P}$ and  $\gamma^{W_P}_{AB} = \gamma^{W_P}$. 

\subsection{General consideration}\label{sec:4add1}

For the isolation of the odd spectra, our basic strategy here is to geometrically identify the networks that have the correspondence
\begin{align}
 {\gamma^{I_P}}\rq{}= -{\gamma^{I_P}} \label{xx}
\end{align}
with respect to a mirror transformation. Since we identically have ${\gamma^{I_P}}\rq{}= {\gamma^{I_P}}$ for an arbitrarily mirror transformation (see Sec. \ref{3q}), we obtain  ${\gamma^{I_P}}=0$ for a network with Eq. (\ref{xx}).

As shown in Fig. 2,  we take the $z$-axis parallel to the direction vector $\hat{\bm{d}}$, and consider the mirror transformation at the $yz$-plane. As in the case of $\hat{\bm{d}}$, The four rank tensors $\Gamma^{I_P}_{ijkl}$ are invariant with this transformation ${\Gamma^{I_P}_{ijkl}}\rq{}=\Gamma^{I_P}_{ijkl}$, and we have
\begin{align}
 {\gamma^{I_P}}\rq{}={{\Gamma^{I_P}_{ijkl}}\rq{}{D_{A}^{ij}}}\rq{} {D_B^{kl}}\rq{}=\Gamma^{I_P}_{ijkl} {D_{A}^{ij}}\rq{} {D_B^{kl}}\rq{}
\end{align}
in comparison to the original one ${\gamma^{I_P}} = \Gamma^{I_P}_{ijkl} {D_{A}^{ij}}{D_B^{kl}}$.  Therefore, we can realize the desired condition ${\gamma^{I_P}}\rq{}=-{\gamma^{I_P}}$ by using two detectors  $A$ and $B$ transformed as  
\begin{align}
\bm{D}_{A}\rq{}=-\bm{D}_{A},~~~\bm{D}_{B}\rq{}=\bm{D}_{B} \label{cond}.
\end{align}

\begin{figure}[t]
\centering
\includegraphics[keepaspectratio, scale=0.2]{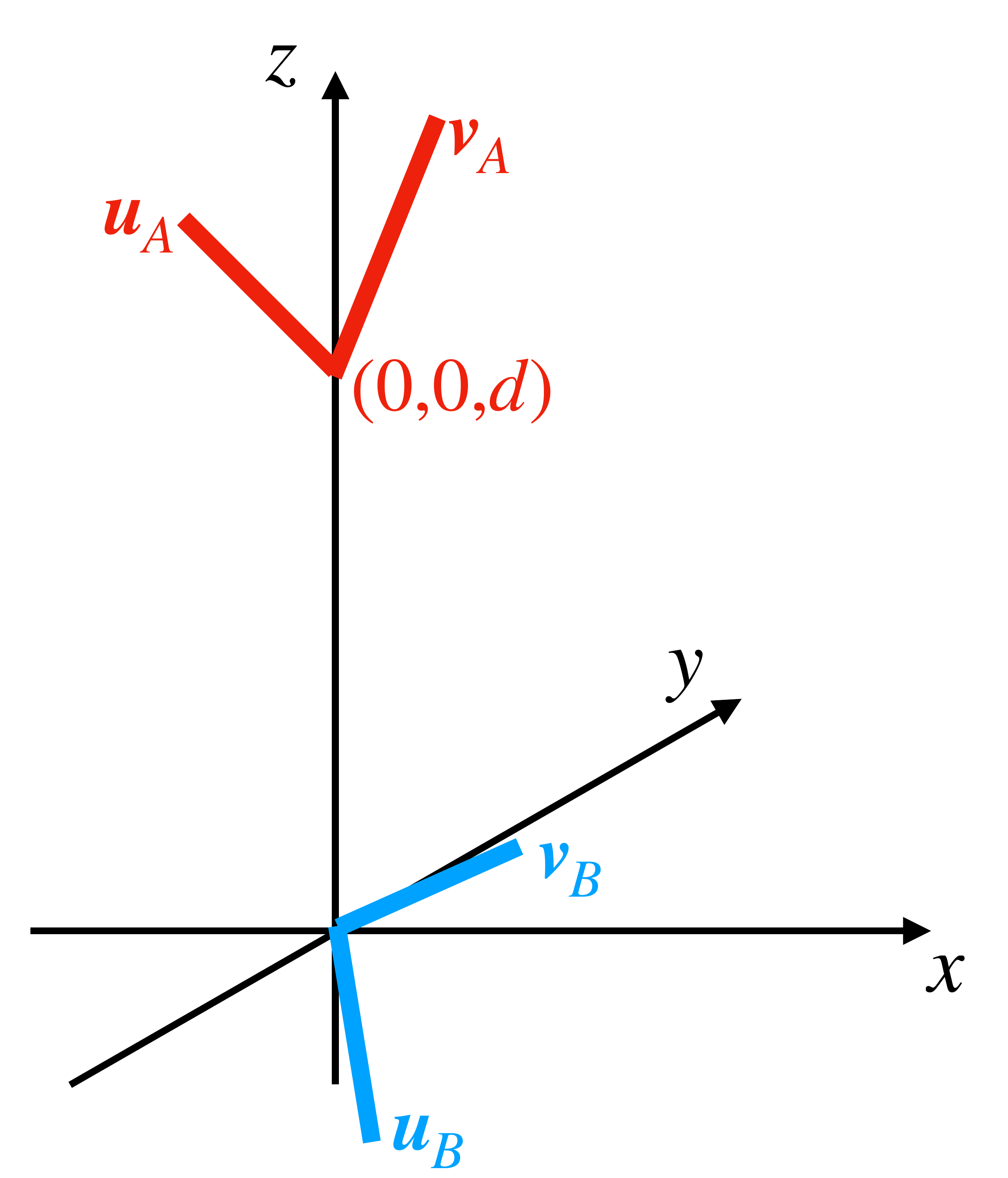}
\caption{The configuration of the detector network under consideration.  We fix the positions of the  two detectors and examine their detector tensors with respect to the reflection (mirror transformation) at the $yz$-plane. }
\label{fig:2}
\end{figure}

Looking back at the arguments so far, one would consider that our networks with Eq. (\ref{cond}) would be just a subset of the networks with the desired property $ {\gamma^{I_P}}=0$.  But, after  various examinations, we deduced that our requirements (\ref{cond}) actually cover  the whole  network geometries  satisfying the identity $ {\gamma^{I_P}}=0$. Hereafter, we assume that this is really the case and call our networks the asymmetric networks. 

\subsection{Detector Tensors}\label{sec:4B}

We now identify detector tensors which are transformed as Eq. (\ref{cond}) with respect to the reflection at the $yz$-plane.

\subsubsection{flipped one}\label{sec:4B1}

\begin{figure}[t]
\centering
\includegraphics[keepaspectratio, scale=0.2]{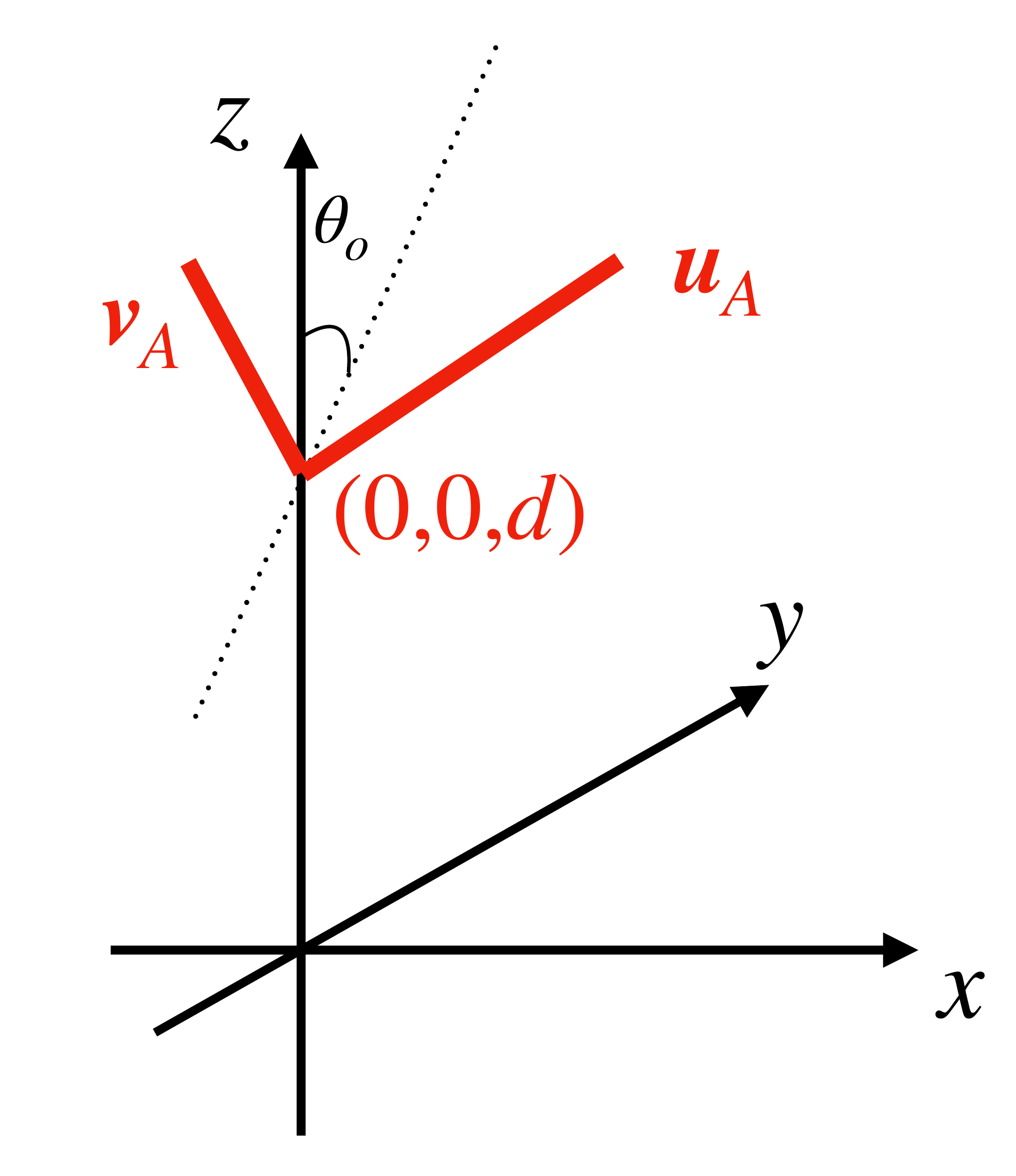}
\caption{The two arm directions for the flipped detector tensor (see Eq. (\ref{eq:61})). The arms are mirror symmetric with respect to the $yz$-plane. We define $\theta_0$ as the angle between their bisector and the $z$-axis.}
\label{fig:3}
\end{figure}

First, we examine the flipped one with $\bm{D}_{A}\rq{}=-\bm{D}_{A}$.  Since a detector tensor is formally given by 
$\bm{D}_{A} = ({\bm{u}_A \otimes \bm{u}_A - \bm{v}_A \otimes \bm{v}_A})/{2}$, we can simply compose the flipped one by using two vectors interchanged as $ \bm{u}_A\rq{}= \bm{v}_A$ and  $ \bm{v}_A\rq{}= \bm{u}_A$. They are mutually mirrored images and  parameterized as  
\begin{align}\label{eq:61}
\begin{aligned}
	\bm{u}_{A} &= \frac{1}{\sqrt{2}}(1,\sin\theta_0,\cos\theta_0)~,  \\
 \bm{v}_A &= \frac{1}{\sqrt{2}}(-1,\sin\theta_0,\cos\theta_0)~
 \end{aligned}
\end{align}
with the angle $\theta_0$ between the bisecting vector and the $z$-axis. (see Fig. \ref{fig:3})

Note that the detector tensor $\bm{D}_{A}$ is given as quadratures of unit vectors, and we can multiply $-1$ to $\bm{u}_A$ and/ or $\bm{v}_A$ in Eq. \eqref{eq:61}, keeping the relation  $\bm{D}_{A}\rq{}=-\bm{D}_{A}$.  In this manner, we can make totally $2\times2$ equivalent pairs of unit vectors. We call this reduplication the ``multiplicity of vector signs".  Correspondingly, the detector tensor with the angle $\theta_0=\alpha+\pi$ is essentially the same as that with $\theta_0=\alpha$.  

\subsubsection{invariant ones}\label{sec:4B2}

\begin{figure}[t]
\centering
\includegraphics[keepaspectratio, scale=0.16]{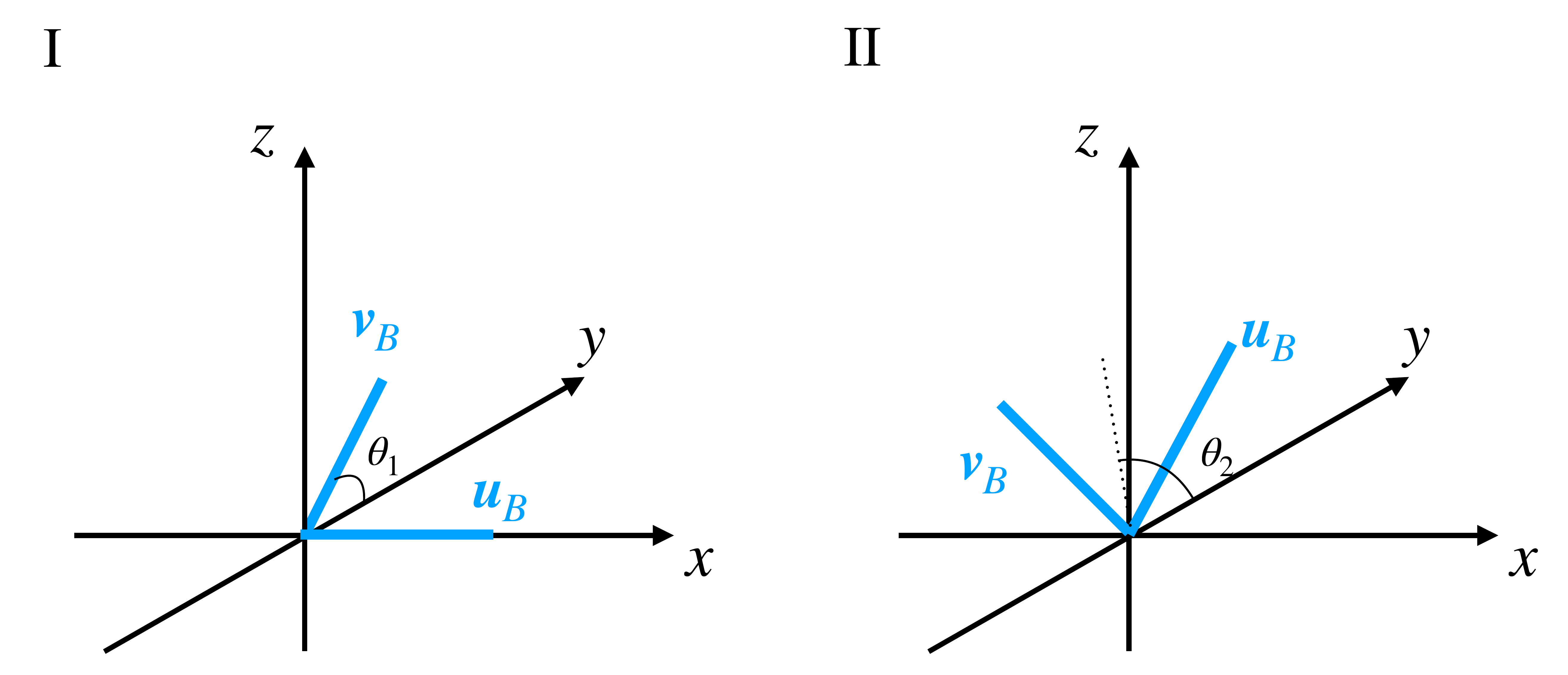}
\caption{(Left) The configuration of the type-I invariant detector tensor (see Eq. (\ref{eq:63})).  The vector $\bm{u}_B$ is on the $x$-axis and  $\bm{v}_B$ is on the $yz$-plane with the angle $\theta_1$ relative to the $y$-axis. (Right)  The configuration of the type-II invariant detector tensor  (see Eq. (\ref{tii})).    The two vectors  are on the $yz$-plane with the bisecting angle $\theta_2$. }
\label{fig:4}
\end{figure}

Next we discuss the detector tensors with $\bm{D}_{B}\rq{}=\bm{D}_{B}$. We can compose it with the unit vectors transformed as 
\begin{align}
\bm{u}_{B}\rq{} &=\pm \bm{u}_{B},~~~\bm{v}_{B}\rq{} =\pm \bm{v}_{B}.
\end{align}
After all, we can find two independent types of solutions.  The first one (type I) is parameterized as
\begin{align}\label{eq:63}
\bm{u}_{B} &= (1,0,0)~,  &\bm{v}_B &= (0,\cos\theta_1,\sin\theta_1)~
\end{align}
with the transformations  $\bm{u}_{B}\rq{} = \bm{u}_{B}$ and $\bm{v}_{B}\rq{} =- \bm{v}_{B}$.  We still have the multiplicity of vector signs, and the detector with the angle $\theta_1=\beta+\pi$ is essentially the same as that with $\theta_1=\beta$. 

The second one (type II) is parameterized as
\begin{align}
\begin{aligned}\label{tii}
\bm{u}_{B} &= (0,\cos\left(\theta_2 - \frac{\pi}{4}\right),\sin\left(\theta_2 - \frac{\pi}{4}\right))~, \\
 \bm{v}_B &= (0,\cos\left(\theta_2 + \frac{\pi}{4}\right),\sin\left(\theta_2 + \frac{\pi}{4}\right))
\end{aligned}
\end{align}
with $\bm{u}_{B}\rq{} =\bm{u}_{B}$ and  $\bm{v}_{B}\rq{} =\bm{v}_{B}$. In  this case, due to the``multiplicity of vector signs", the detector with $\theta_2=\gamma+\pi/2$ is essentially the same as $\theta_2=\gamma$.  We should notice that the phase offset $\pi/2$ is different from $\pi$ associated with  $\theta_0$ and $\theta_1$ in Eqs. \eqref{eq:61} and \eqref{eq:63}.

\subsection{Parity Odd ORFs}\label{sec:4C}

\begin{figure}[t]
\centering
\includegraphics[keepaspectratio, scale=0.15]{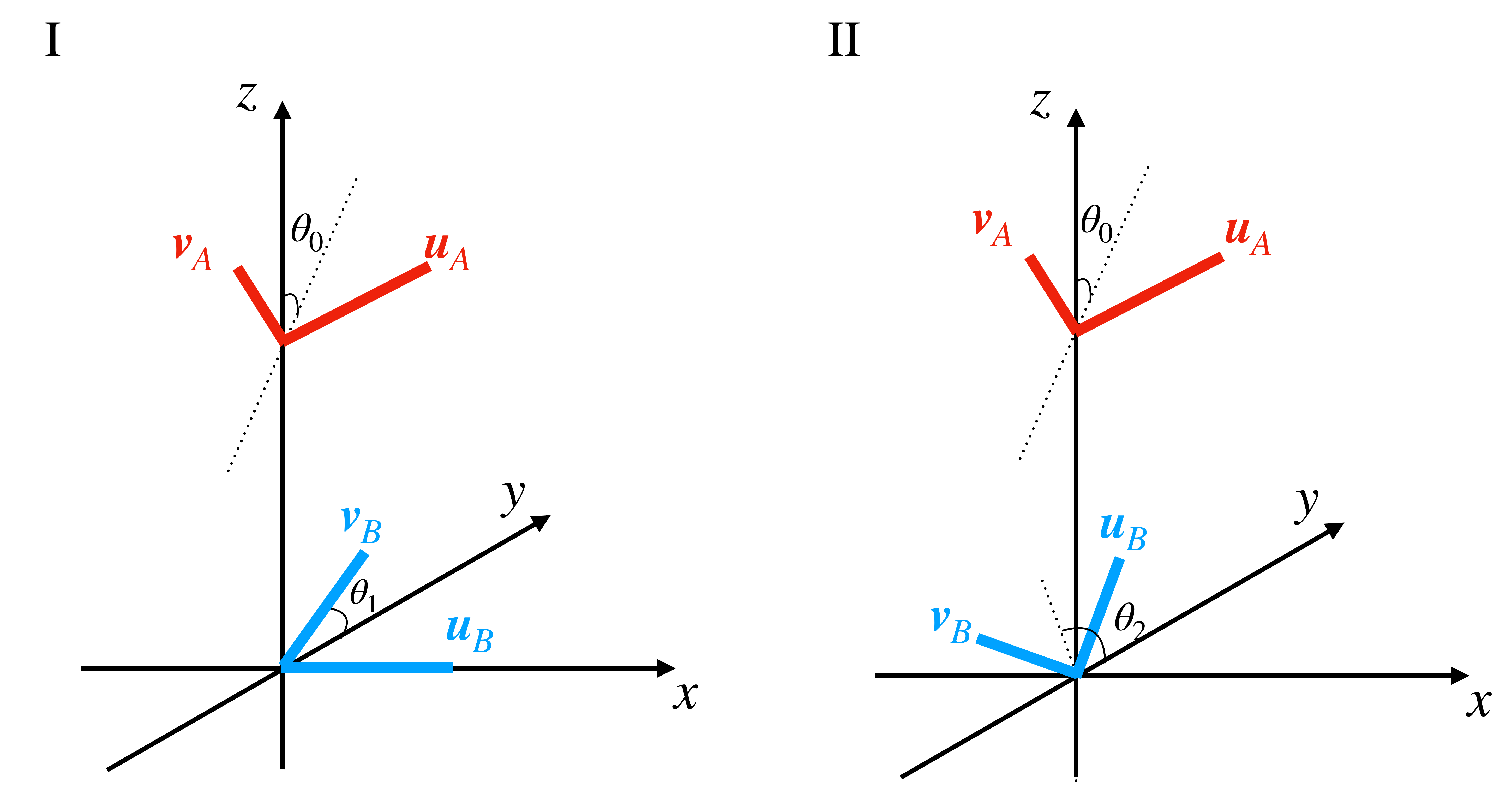}
\caption{(Left)   The type-I network formed by combining one flipped detector and one type-I invariant detector. This network is characterized by two angles $\theta_0$ and $\theta_1$ (see Figs. \ref{fig:3} and \ref{fig:4} for their definitions).  (Right) The type-II network formed by combining one flipped detector and one type-II invariant detector. This network is characterized by the angular parameters $\theta_0$ and $\theta_2$ (see Figs. \ref{fig:3} and \ref{fig:4} for their definitions).  }
\label{fig:5}
\end{figure}

 As mentioned in Sec. \ref{sec:4add1}, an asymmetric network is insensitive to the parity even spectra and allows us to exclusively  probe  the parity odd spectra.  Such a network can be formed by combining one invariant detector and one flipped detector, as shown in Eq. \eqref{cond}.   Following the classification of the invariant detector, we divide the asymmetric networks into the types I and II (see Fig. 5). 

We now evaluate the parity odd ORFs $\gamma^{W_P}$ ($P=T$ and $V$) for the asymmetric networks. We deal with the two types separately in the following subsections. Based on the interest in realizing good sensitivities, we examine the maximums of the absolute values $|\gamma^{W_p}|$.   From our experience in mathematical analysis, we expect that the maximum values  will be obtained for highly symmetric network geometries.

\subsubsection{type {\rm I} network}\label{sec:4C1}

We first derive the analytic expressions of the functions $\gamma^{W_p}$ for the type I network.  With  Eqs. \eqref{eq:61} and \eqref{eq:63} for the orientations of the arms, we obtain the detector tensors \eqref{eq:19} as 
\begin{align}\label{eq:65x}
	\bm{D}_A &= \frac{1}{2} \left(
	\begin{array}{ccc}
		0 & \sin\theta_0 & \cos\theta_0\\
		\sin\theta_0 & 0 & 0\\
		\cos\theta_0 & 0 & 0
	\end{array}
	\right)~,\\
	\label{eq:66x}
	\bm{D}^{\rm I}_B &= \frac{1}{2} \left(
	\begin{array}{ccc}
		1& 0 & 0\\
		0 & -\cos^2\theta_1 & -\frac{1}{2} \sin 2\theta_1\\
		0 & -\frac{1}{2} \sin 2\theta_1 & -\sin^2\theta_1
	\end{array}
	\right)~.
\end{align}
Then, using Eqs. \eqref{eq:49} and \eqref{eq:50}, we have
\begin{align}\label{eq:64}
\begin{aligned}
	\gamma^{W_T}_{\rm I}(y;\theta_0,\theta_1) = &-\frac{1}{8}(3+\cos2\theta_1) \sin\theta_0 \left(4 j_1 - j_3\right)\\
	&\add{-}  \frac{1}{2}\sin2\theta_1 \cos\theta_0 (j_1 + j_3)~,
\end{aligned}\\
	\label{eq:65}
	\begin{aligned}
	\gamma^{W_V}_{\rm I}(y;\theta_0,\theta_1) = &-  \frac{1}{4}(3 + \cos2\theta_1 )\sin\theta_0 (j_1 + j_3)\\
	&\add{-} \frac{1}{4} \sin2\theta_1 \cos\theta_0 \left(j_1 - 4j_3\right)~.
\end{aligned}
\end{align}

We comment on basic properties of these expressions.  First, there exist the periodicities $\theta_0 \to \theta_0 +2 \pi$ and $\theta_1 \to \theta_1 + \pi$. In addition, for the absolute values of these functions, the periodicities reduce to $\theta_0 \to \theta_0 + \pi $ and $\theta_1 \to \theta_1 + \pi$, as expected from  the  ``multiplicity of vector signs"  pointed out in Sec. \ref{sec:3B}.

\begin{figure}
\centering
\includegraphics[keepaspectratio, scale=0.15]{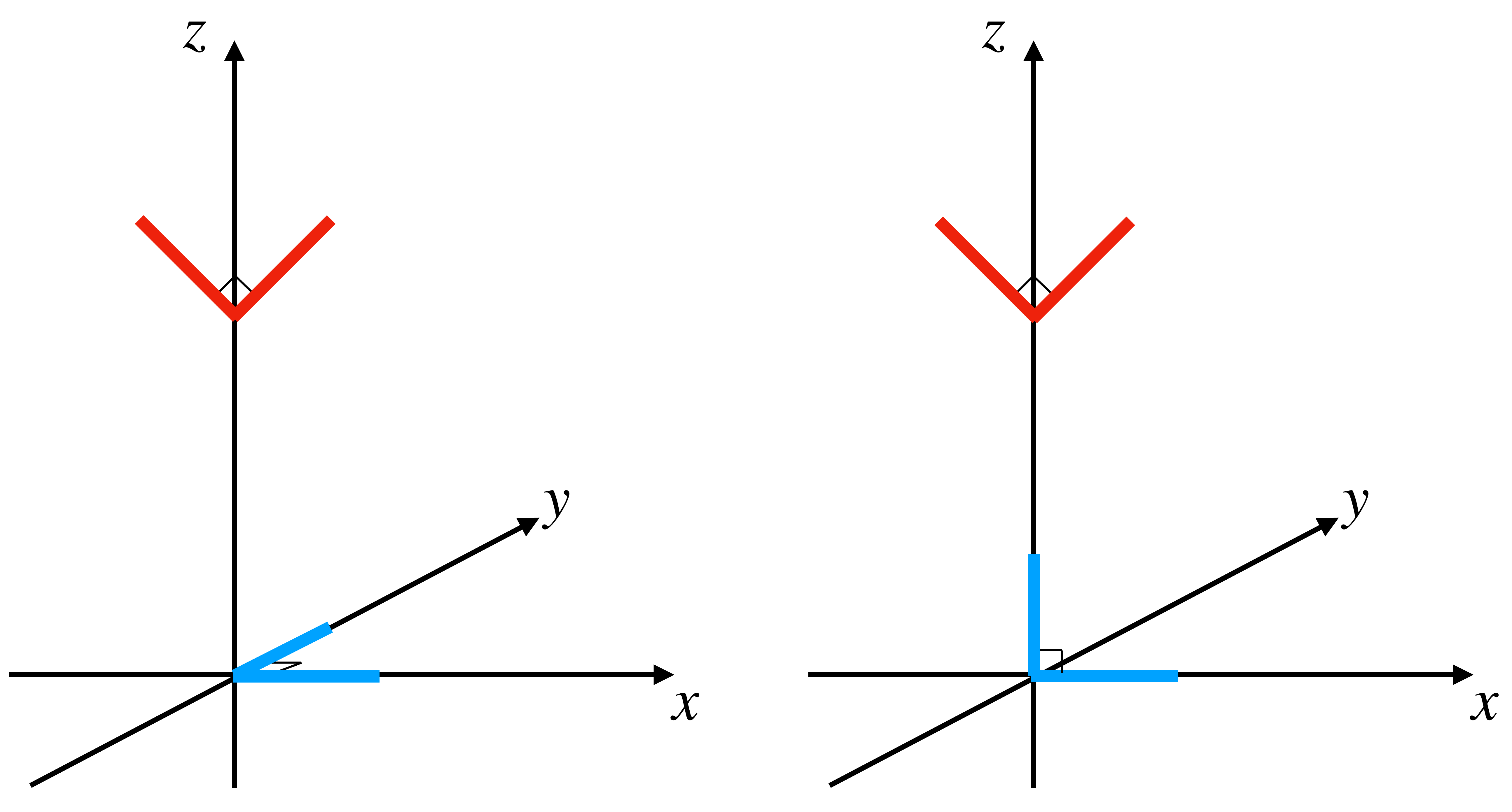}
\caption{The highly symmetric configurations of the  type-I network resulting in Eq. (\ref{zero}).  All the five ORFs vanish for these networks. (Left) The configuration with the angular parameters  $(\theta_0,\theta_1) = (0,0)$. The upper detector (red lines) is on the $xz$-plane and the lower detector (blue lines) is on the $xy$-plane. (Right) The configuration with the angular parameters  $(\theta_0,\theta_1) = (0,\pi/2 )$. Both upper detector (red lines) and lower detector (blue lines) are on the $xz$-plane. }
\label{fig:6}
\end{figure}

Secondly, the two ORFs identically become 
\begin{align}\label{zero}
\gamma^{W_V}_{\rm I}=\gamma^{W_T}_{\rm I}=0
\end{align}
 at $(\theta_0,\theta_1) = (0,0)$ and $(0,\pi/2)$  (see Fig. \ref{fig:6} for the corresponding configurations).  In these highly symmetric configurations, the network is parity even with respect to the reflection at the $xz$-plane.   Then,  following an argument similar to Sec. \ref{sec:4add1}, we can readily confirm  Eq. \eqref{zero} (see also \cite{Seto:2008sr} for an earlier discussion on the tensor modes).   We should notice that, the networks in Fig. 6 are still parity odd for the reflection at the $yz$-plane and all of the five ORFs  vanish in Eq. \eqref{c18}.

We can also find  that the  ORFs \eqref{eq:64} and \eqref{eq:65}  are invariant under the following transformation
\begin{align}
\begin{cases} \label{inv}
	\theta_0 \to \pi -\theta_0~,\\
	\theta_1 \to \pi - \theta_1~.
\end{cases}
\end{align}
Geometrically, this corresponds to taking the reflection at the $xz$-plane and subsequently interchanging two arms of the upper detector (detector A) in the left panel in Fig. 5. The overall signs of the ORFs are changed twice and we recover the original forms.  

\begin{figure*}[t]
\centering
\includegraphics[keepaspectratio, scale=0.5]{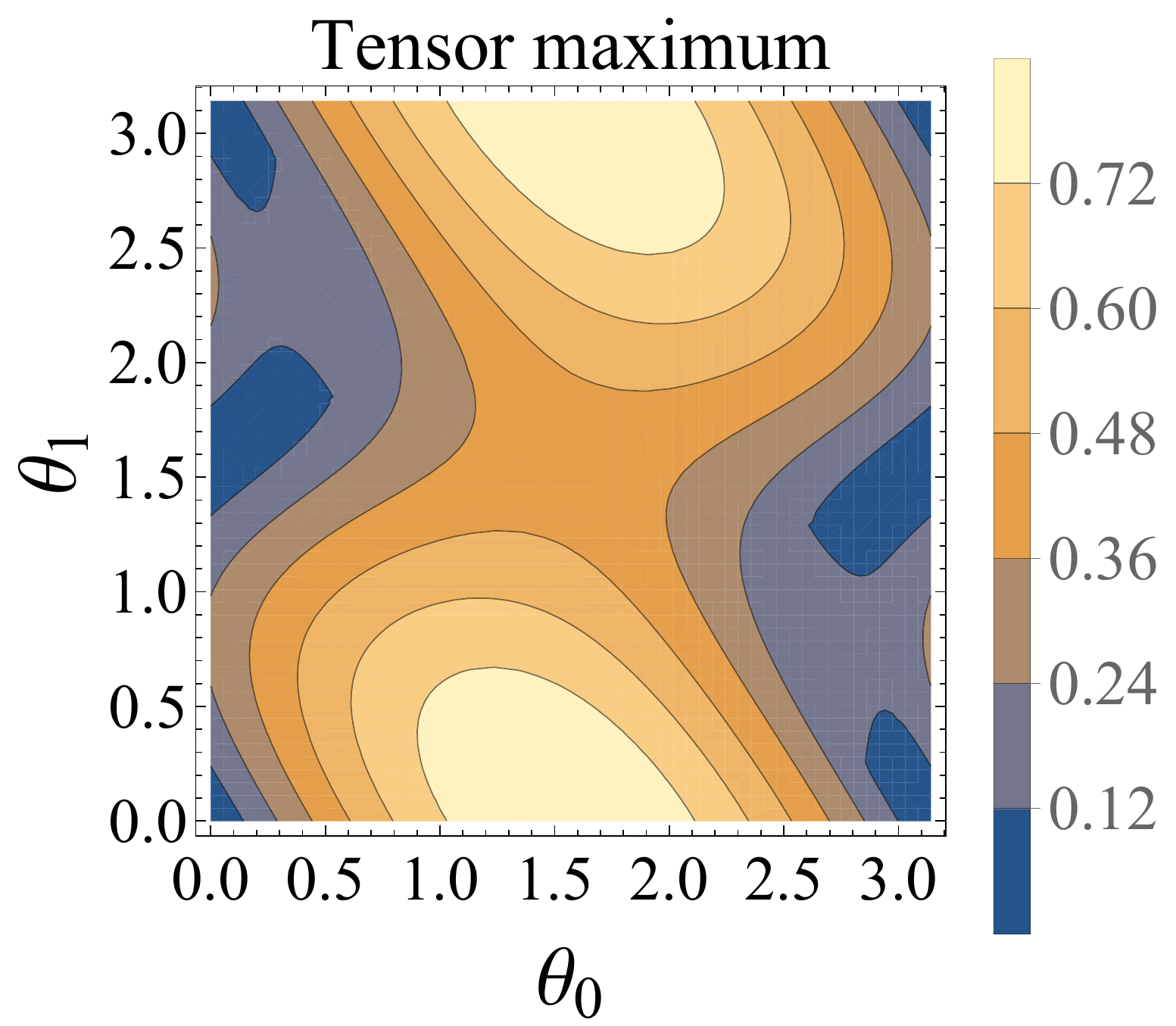}
\includegraphics[keepaspectratio, scale=0.5]{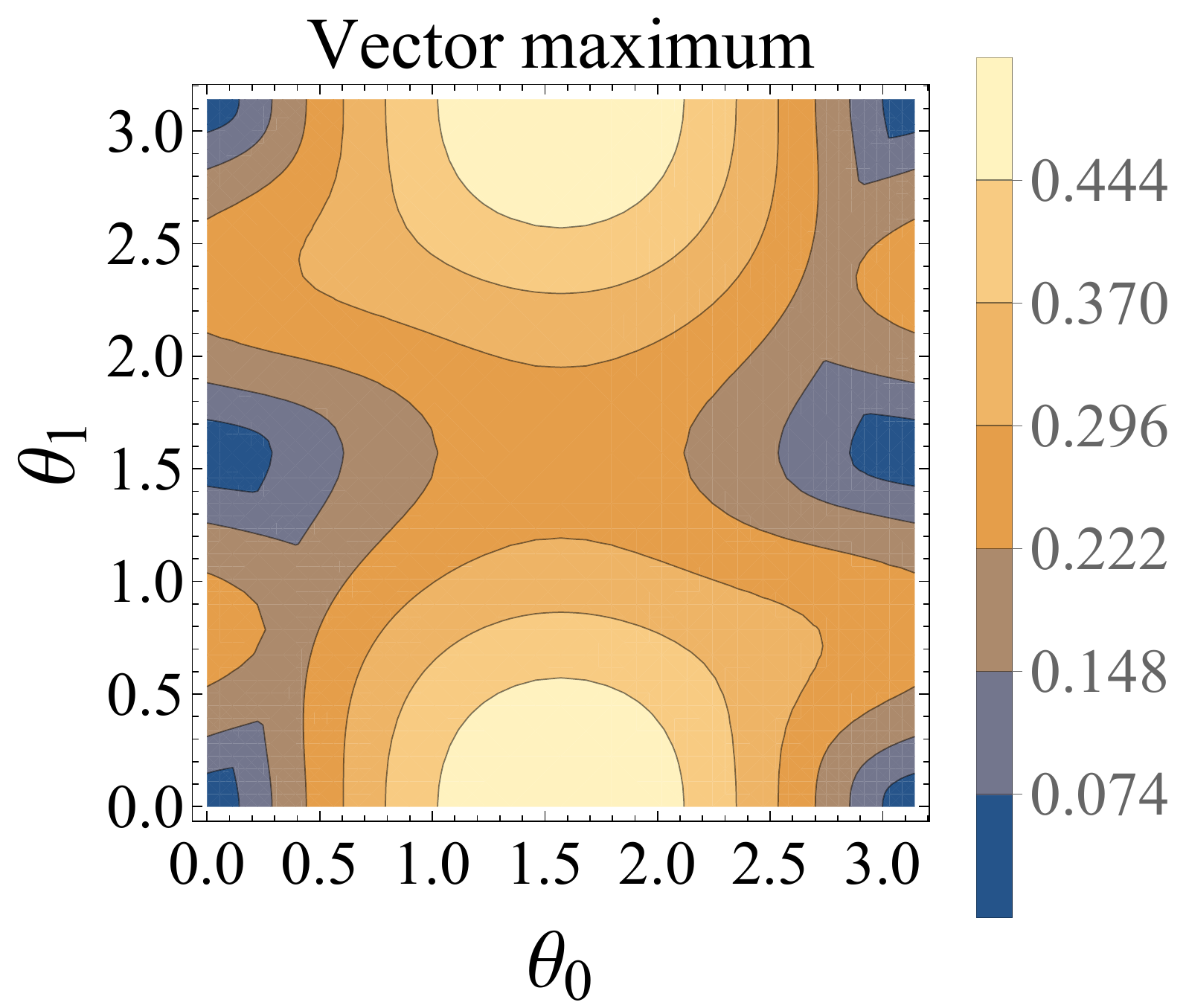}
\caption{(Left) The contour plot of the function $F_{\rm I}^{W_T}(\theta_0,\theta_1)$ defined in Eq. (\ref{eq:74}). The minimum value is 0 and the maximum value is $0.840$. (Right) The contour plot of the function $F_{\rm I}^{W_V}(\theta_0,\theta_1)$ ranging from 0 to 0.520.}
\label{fig:7}
\end{figure*}

\begin{figure}[t]
\centering
\includegraphics[keepaspectratio, scale=0.15]{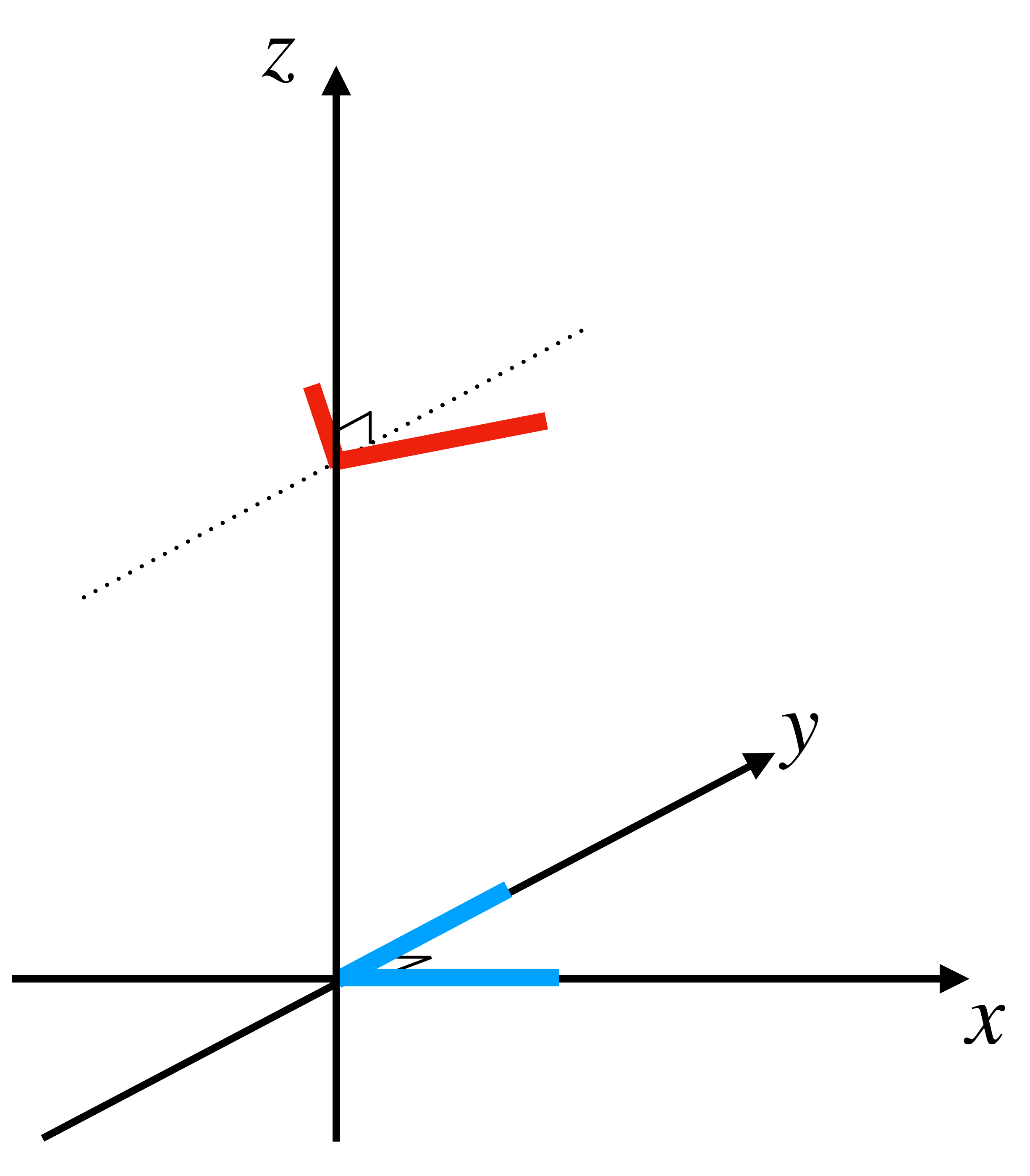}
\caption{Highly symmetric network configuration with $(\theta_0,\theta_1) = (\pi/2,0)$ corresponding to the maximums in Fig. \ref{fig:7}. Both upper detector (red lines) and lower detector (blue lines) are parallel to the $xy$-plane.  This network is parity odd for the reflections  at the $xz$- and $yz$-planes.}
\label{fig:8}
\end{figure}

Now we examine the network configurations that realize large absolute  values  $|\gamma^{W_P}_{\rm I}(y,\theta_0,\theta_1)|$ for the  two odd modes $P=T$ and $V$. 
To begin with, for given angular variables $(\theta_0,\theta_1)$ and  an index $P$, we numerically identify the parameters $y=y^P_{m,{\rm I}}(\theta_0,\theta_1)$ that maximize the functions  $|\gamma^{W_P}_{\rm I}(y,\theta_0,\theta_1)|$. We then define the resultant maximum values by
\begin{align}\label{eq:74}
	F_{\rm I}^P(\theta_0,\theta_1) \equiv |\gamma_{\rm I}^{W_P}(y^P_{m,{\rm I}}(\theta_0,\theta_1);\theta_0,\theta_1)| ~.
\end{align}
In Fig. \ref{fig:7},  we show their contour plots. We can easily confirm the properties mentioned earlier, such as the  periodicities, the invariance under transformation \eqref{inv}, and $\gamma^{W_T}_{\rm I} = \gamma^{W_V}_{\rm I} = 0$  at $(\theta_0,\theta_1) = (0,0)$ and $(0,\pi/2)$. 

The global maximum values of the functions $F_{\rm I}^P$ are commonly at $(\theta_0,\theta_1)=(\pi/2,0)$ with the combinations
\begin{align}
	F_{\rm I}^T(\pi/2,0) \simeq& 0.840~ & {\rm }~~y_{ m,T}(\pi/2,0) &\simeq 2.000~, \label{m1t}\\
	F_{\rm I}^V(\pi/2,0) \simeq& 0.520~ & {\rm }~~y_{ m,V}(\pi/2,0) &\simeq 2.501~.\label{m1v}
\end{align}
 In Fig. \ref{fig:8},  we show the network configuration with $(\theta_0,\theta_1)=(\pi/2,0)$.  We can find that this network is parity odd with respect to both the $xz$- and $yz$-planes. This also supports our naive expectation that sensitivity is maximized for a highly symmetric configuration.
 
 In comparison, for the even parity modes, the maximum values of the ORFs are simply given by $\gamma^{I_T}(y=0)=\gamma^{I_V}(y=0)=\gamma^{I_S}(y=0)=1$ for two co-aligned detectors (with $\gamma^{W_T}(y)=\gamma^{W_V}(y)=0$). 

\subsubsection{type {\rm II} network}\label{sec:3C2}

Next, we consider the type II network. The analysis here is parallel to the previous subsubsection for the type I network. The geometrical difference between the two types is the orientation of the  invariant detector ($B$ in Fig. \ref{fig:5}). Its  detector arms are given by Eq. \eqref{tii} with the detector tensor 
\begin{align}
	\bm{D}^{\rm II}_B &= \frac{1}{2} \left(
	\begin{array}{ccc}
		0& 0 & 0\\
		0 & \sin2\theta_2 & -\cos 2\theta_2\\
		0 & -\cos 2\theta_2 & -\sin2\theta_2
	\end{array}
	\right)~.
\end{align}
Then we obtain the analytic expressions for the ORFs as
\begin{align}
\begin{aligned}
	\gamma^{W_T}_{\rm II} = & \add{-}\frac{1}{4} \sin2\theta_2 \sin\theta_0 \left(-4 j_1+j_3\right)\\
	& -  \cos2\theta_2 \cos\theta_0 (j_1 + j_3)~, \label{g2t}
\end{aligned}\\
\begin{aligned}
	\gamma^{\add{W_V}}_{\rm II} = &\add{+}  \frac{1}{2} \sin2\theta_2 \sin\theta_0 (j_1 + j_3)\\
	&+\frac{1}{2} \cos2\theta_2 \cos\theta_0  \left( -j_1 + 4j_3\right)~.\label{g2v}
\end{aligned}
\end{align}

The basic properties of these functions are  similar to those already mentioned for the type I network. More specifically,  we have the identical periodicities $\theta_0 \to \theta_0 + 2\pi~$ and $\theta_2 \to \theta_2 + \pi$. In  contrast, for their absolute values, the periods  result in $\theta_0 \to \theta_0 + \pi$ and $\theta_2 \to \theta_2 + \frac{\pi}{2}$ with the factor 2 difference for the latter (reflecting the ``multiplicity of vector signs'' noted in Sec. \ref{sec:3B}).

\begin{figure}[t]
\centering
\includegraphics[keepaspectratio, scale=0.15]{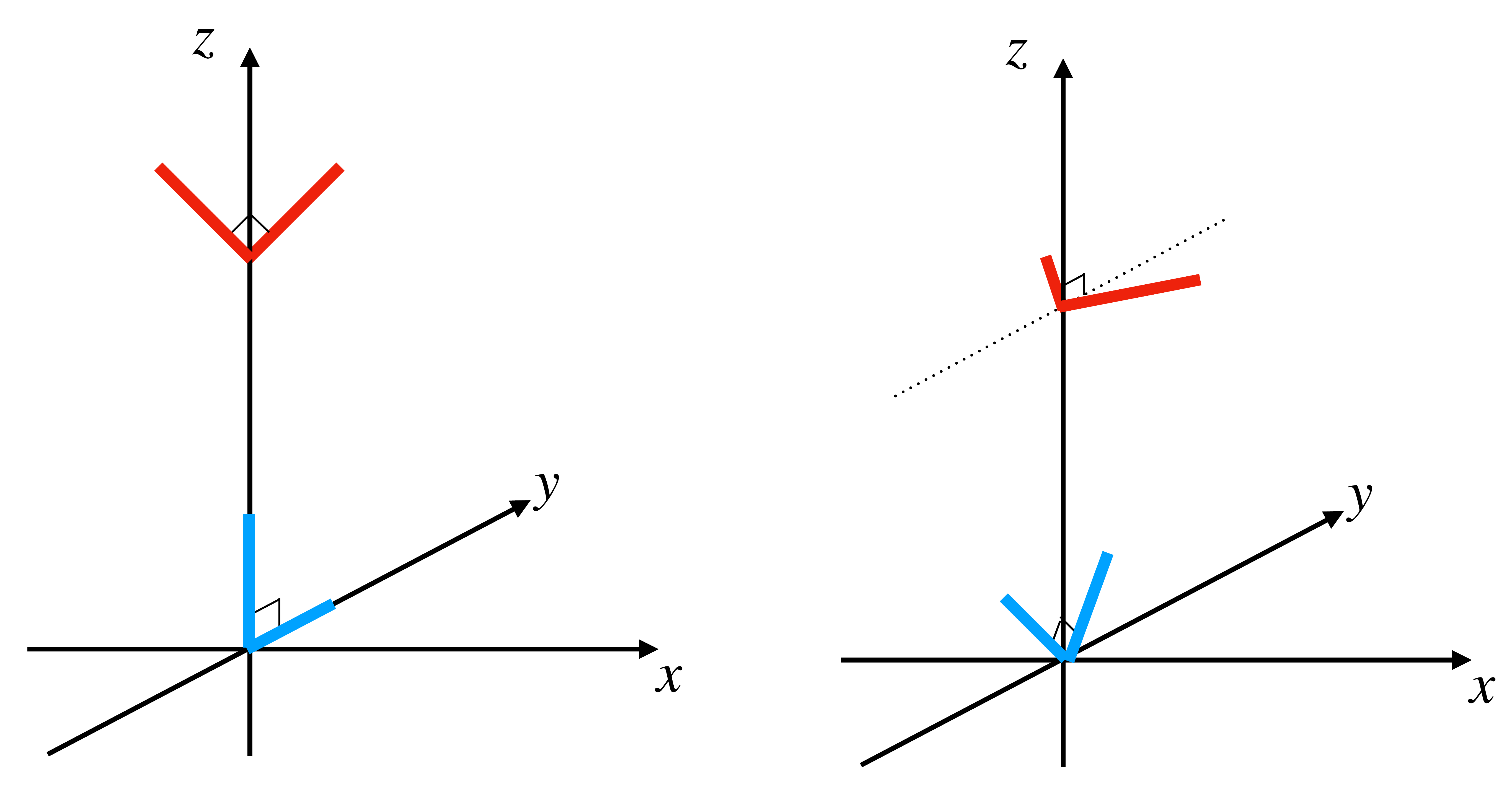}
\caption{The highly symmetric configurations of the  type-II network resulting in Eq. \eqref{eq:86}. All the five ORFs vanish for these networks. (Left) The configuration with the angular parameters  $(\theta_0,\theta_2) = (0,\pi/4)$. The upper detector (red lines) is on the $xz$-plane and the lower detector (blue lines) is on the $yz$-plane. (Right) The configuration with the angular parameters  $(\theta_0,\theta_2) = (\pi/2,\pi/2 )$. The upper detector (red lines) is parallel to the $xy$-plane and the lower detector (blue lines) is on the $yz$-plane with bisector of the two arms on the $z$ axis.}
\label{fig:9}
\end{figure}

In addition, we have the identities 
\begin{align}\label{eq:86}
	\gamma_{\rm II}^{W_T} = \gamma_{\rm II}^{W_V} = 0~,
\end{align}
at $(\theta_0,\theta_2) = (0,\pi/4)$ and $(\pi/2,0)$ (see Fig. \ref{fig:9} for the corresponding configurations), again reflecting the parity evenness with respect to the reflection at the $xz$-plane. 

In contrast to the invariance of the type I network under transformation \eqref{inv}, the functions  (\ref{g2t}) and (\ref{g2v}) change the overall signs under the simultaneous transformations
\begin{align}
\begin{cases}
	\theta_0 \to \pi -\theta_0~,\\
	\theta_2 \to \pi - \theta_2~.
\end{cases}
\end{align}
The geometrical interpretation is almost the same as the previous case, except for the additional interchange of the two arms of the lower detector B in the right panel of Fig. \ref{fig:5}.  This additional operation results in the extra minus signs, compared to the case with Eq. (\ref{inv}).

\begin{figure*}[t]
\centering
\includegraphics[keepaspectratio, scale=0.5]{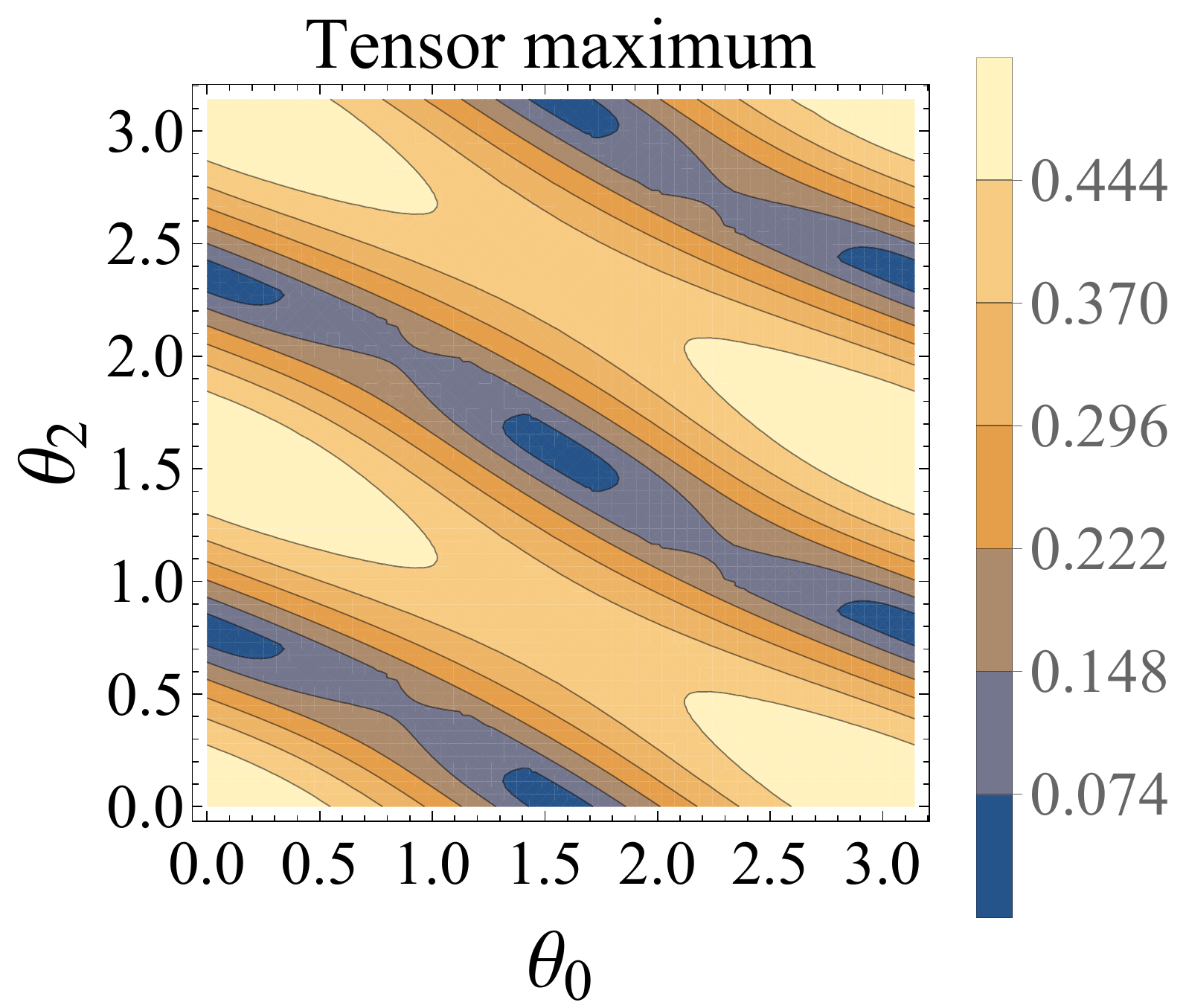}
\includegraphics[keepaspectratio, scale=0.5]{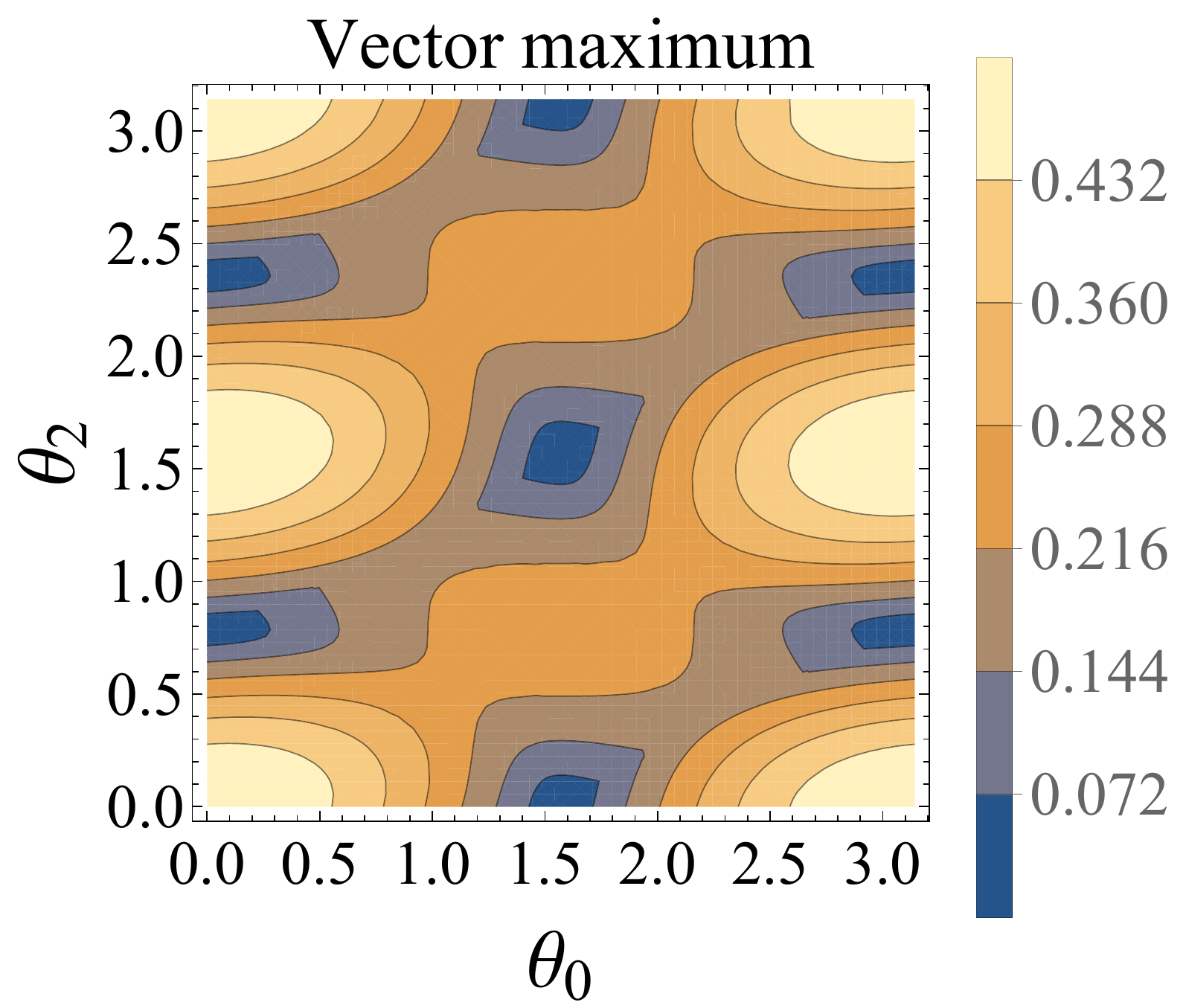}
\caption{(Left) The contour plot of the function $F_{\rm II}^{W_T}(\theta_0,\theta_2)$ defined in Eq. (\ref{eq:84}). The minimum value is 0 and the maximum value is $0.520$. (Right) The contour plot of the function $F_{\rm II}^{W_V}(\theta_0,\theta_2)$ ranging from 0 to 0.508.}
\label{fig:10}
\end{figure*}

Now we study the network configuration which maximizes the absolute values  $|\gamma^{W_P}_{\rm II}|$.  Following the same strategy as before, we define the two functions ($P=T$ and $V$)
\begin{align}\label{eq:84}
	F_{\rm II}^P(\theta_0,\theta_2) \equiv |\gamma_{\rm II}^{W_P}(y^P_{m,{\rm II}}(\theta_0,\theta_2);\theta_0,\theta_2)|~,
\end{align}
where $y^P_{m,{\rm II}}(\theta_0,\theta_2)$ maximize $\gamma^{W_P}_{\rm II}(y;\theta_0,\theta_2)$ for fixed $(\theta_0,\theta_2)$. In Fig. \ref{fig:10}, we show the contour plots of $F^{T}_{\rm II}$ and $F^{V}_{\rm II}$. We can confirm the basic properties of the ORFs mentioned earlier.

\begin{figure}[t]
\centering
\includegraphics[keepaspectratio, scale=0.15]{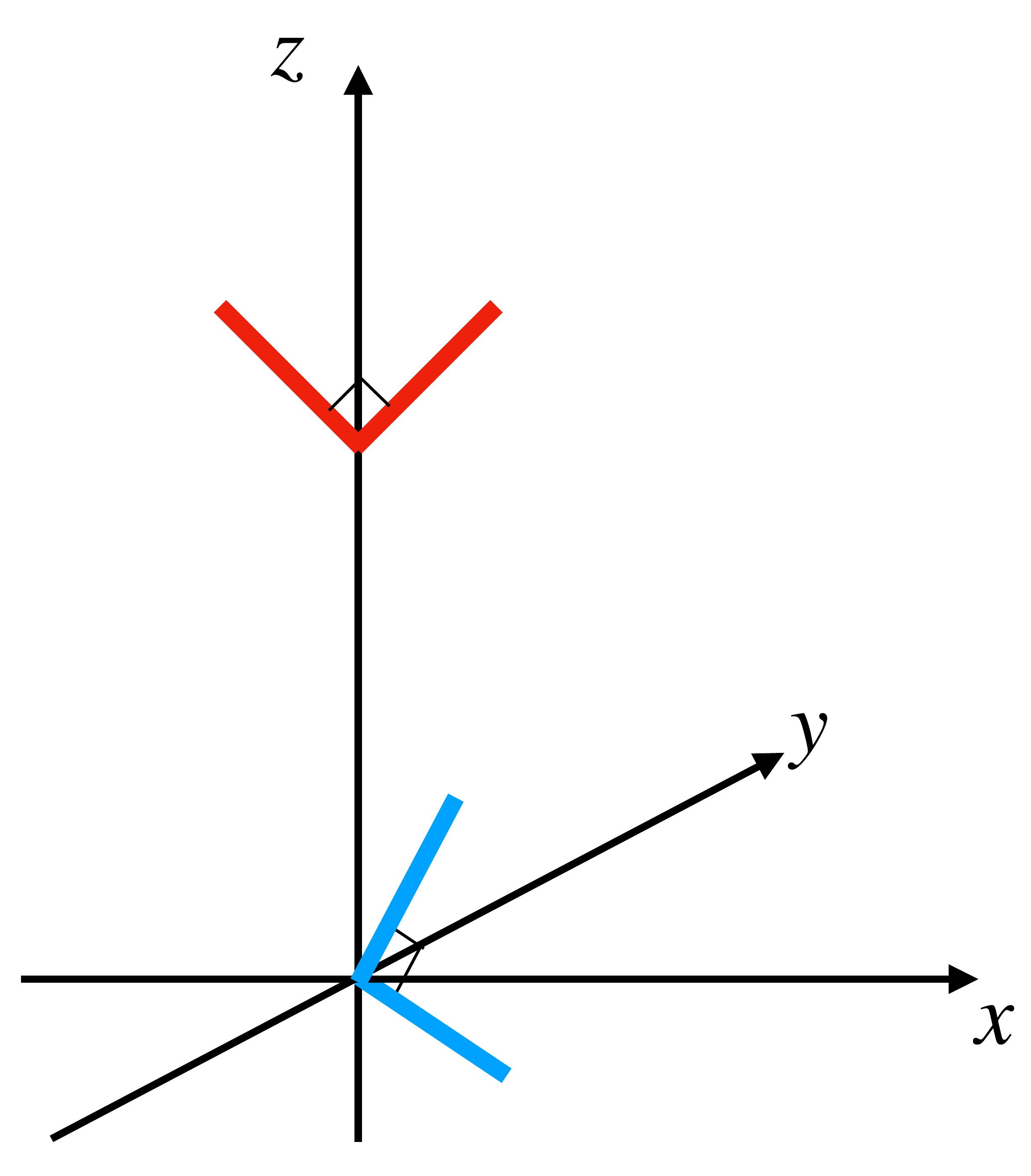}
\caption{Highly symmetric network configuration with $(\theta_0,\theta_2) = (0,0)$ corresponding to the maximums in Fig. \ref{fig:10}. The upper detector (red lines) is on the $xz$-plane and lower detector (blue lines) is on the $yz$-plane with bisector of two arms on the $y$ axis.  This network is parity odd for the reflections  at the $xz$- and $yz$-planes.}
\label{fig:11}
\end{figure}

The global maximums of the type II detectors are realized at $(\theta_0,\theta_2) = (0,0)$ with
\begin{align}
	F_{\rm II}^T(0,0) &\sim 0.520~,  & y_{m,{\rm II}}^T &\sim 2.501~,\\
	F_{\rm II}^V(0,0) &\sim 0.508~,  & y_{m,{\rm II}}^V &\sim 4.921~.
\end{align}
Fig. \ref{fig:11} shows the corresponding configuration with $(\theta_0,\theta_2) = (0,0)$.  Again, this configuration is parity odd for the reflections  both at the $yz$- and $xz$- planes. However, the maximum values are smaller than Eqs. \eqref{m1t} and \eqref{m1v}. 

\subsection{Relation to the ground-based networks}

\begin{figure}[t]
\centering
\includegraphics[keepaspectratio, scale=0.2]{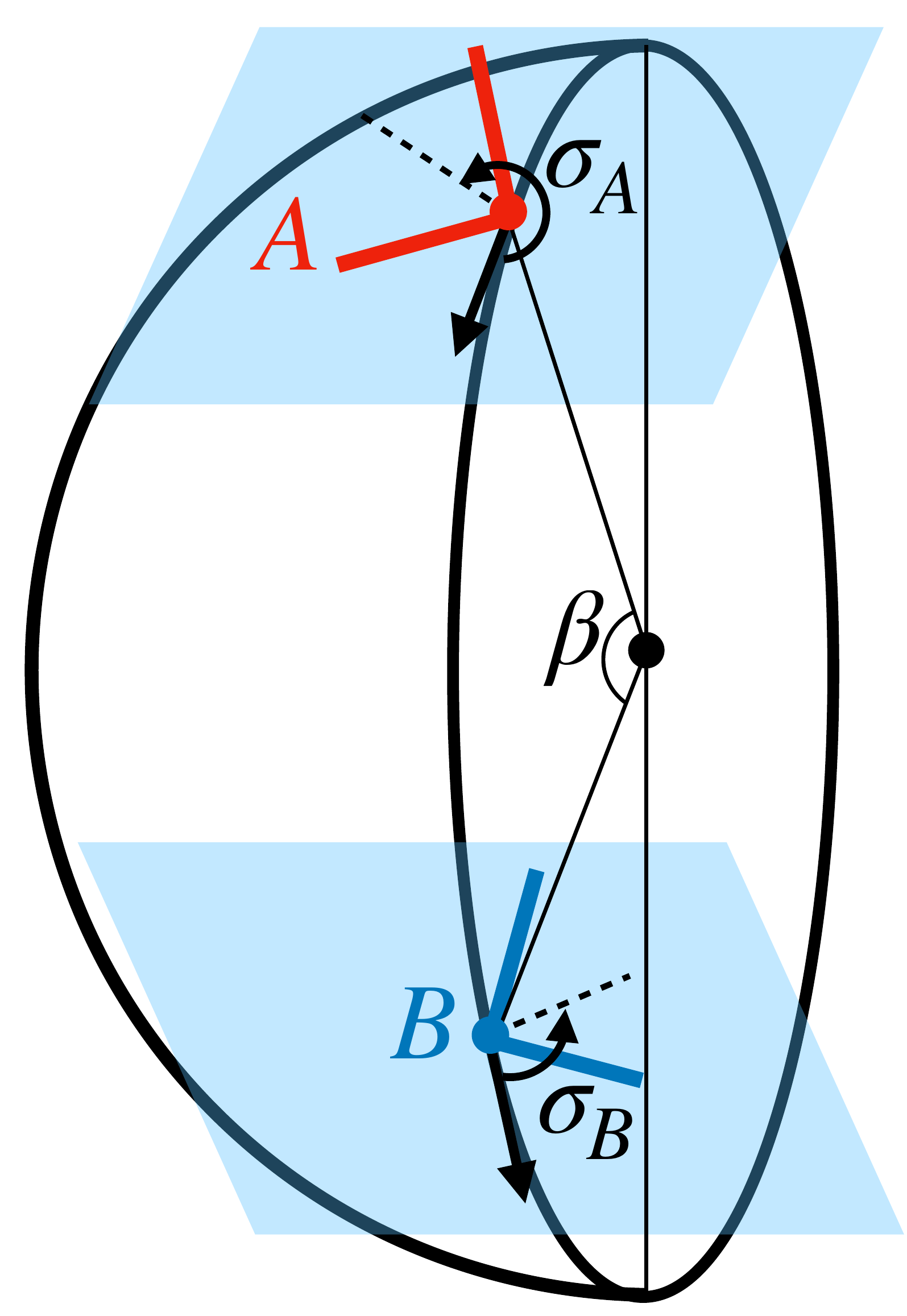}
\caption{The relative configuration of   two ground based detectors A and B. The two detectors are on the blue planes which are tangential to the Earth sphere. The angle $\beta$ shows the separation of two detectors, measured from the center of the Earth. The two angles $\sigma_A$ and $\sigma_B$ describe the orientation of the bisectors of the two detector arms (dotted lines) in a counterclockwise manner, relative to the great circle connecting the two detectors. The angular parameters $\Delta$ and $\delta$ are defined by $\Delta \equiv (\sigma_A + \sigma_B)/2)$ and $\delta \equiv (\sigma_A - \sigma_B)/2$.}    
\label{fig:12}
\end{figure}

Until now, we have discussed the  ORFs   for  highly symmetric configurations purely from geometrical viewpoints. Here, for the odd parity ORFs, we point out the connection of the type I network to a  network composed of two ground based detectors that are tangential to the Earth sphere. 
 
As presented in appendix B, we have the expression for the latter as
 \begin{align}\label{eq:89}
	\gamma^{W_P}_{} &= \Xi^{P}(y,\beta) \sin4 \Delta~, & (P &= T,V)
\end{align}
with the two angular parameters $\beta$ and $\Delta$ (see Fig. \ref{fig:12} for their definitions).  The first one $\beta$ is the angular separation between two detectors measured from the center of the Earth.  The second one $\Delta$ is determined by the orientations of two detectors as
\begin{align}
	\Delta = \frac{\sigma_A + \sigma_B}{2}~.
\end{align}
By taking $|\sin4\Delta|=1$ we can maximize  $|\gamma^{W_P}_{} |= |\Xi^{P}(y,\beta)| $. 

In fact, the function $|\Xi^{P}(y,\beta)| $ is identical to the odd parity ORFs  $|\gamma_{\rm I}^{W_p}|$ given in Eqs.  (\ref{eq:64}) and (\ref{eq:65}) for the type I network,  after tuning its angular parameters $\theta_0$ and $\theta_1$.  
More specifically, we impose the relation 
\begin{align}\label{eq:add2}
	|\theta_1-\theta_0| = \pi/2~,
\end{align} 
corresponding  to the condition that two detectors A and B in Fig. 5 are tangential to a sphere.  Under the relation  (\ref{eq:add2}), the separation angle $\beta$ is given by 
\begin{align}
	\beta &= 
	\begin{cases}
	2\theta_0~, & (0 < \theta_0 < \frac{\pi}{2})~,\\
	2\pi -2\theta_0~, & (\frac{\pi}{2} < \theta_0 < \pi)~,
	\end{cases}
\end{align}
and we obtain  $|\Xi^{P}(y,\beta)|=|\gamma_{\rm I}^{W_p}(y,\theta_0,\theta_1)| $  ($P=T,V$).

\section{Summary and discussion}\label{sec:5}

In this paper, we studied the correlation analysis  for detecting various polarization modes of a stationary and isotropic stochastic gravitational wave background.  We pointed out that, as long as the low frequency approximation is valid, we can probe the five spectra $I_T, I_V, I_S, W_T,$ and $W_V$. The three spectra $I_P (P=T,V,S)$ represent the  intensities of the tensor, vector, and scalar modes. While the remaining two spectra  $W_P (P=T,V)$ show the chiral asymmetries of the tensor and vector modes. Other correlations, such as the tensor-vector pair start from the higher multipole components and thus do not contribute to the monopole components.  

When performing the correlation analysis, the ORFs play  key roles and characterize the correlated responses of detectors to the five spectra. In this paper, we newly derived the function $\gamma^{W_V}$ for the parity odd vector modes and completed all the ORFs required for generally analyzing polarization states of an isotropic background.  For its derivation,  we applied a systematic  method,  explicitly respecting the rotational symmetry (SO(3)) of the system.  We also paid special attention to the parity transformation of the system.  These help us to understand the symmetrical structure of the ORFs and their building blocks.  

Furthermore, we examine two detector networks with respect to reflection transformations which are  closely related to the parity transformation.  For a reflection, the ORFs have the same parity signatures as the corresponding spectra. This property allows us to easily design detector networks that are sensitive to either  even or odd spectra of a background. Such networks are particularly interesting for clearly isolating the two parity.  In Fig. \ref{fig:5}, we illustrated the two types of network geometries that are insensitive to the even ones. 

Then we  examined the odd ORFs $\gamma^{W_P} (P=T,V)$ specifically for the two geometrical types.  When we tune the networks to maximize the sensitivity, they take a highly symmetric configuration which is parity odd  simultaneously to the two different  planes. In contrast, for other highly symmetric configurations with angular parameters $(\theta_0,\theta_1)= (0,0),(0,\pi/2)$ and $(\theta_0,\theta_2) = (0,\pi/4),(\pi/2,\pi/2)$ (see Figs. \ref{fig:6} and \ref{fig:9}),  we identically have $\gamma^{I_P}=0$ $(P = T,V,S)$ and $\gamma^{W_P}=0$ $(P=T,V)$, and thus the detector networks become blind to all the five monopole components.

In this paper, we have concentrated on the basic properties of the ORFs. Now we comment on potential applications of our results. One of the immediate studies would be the prospect for the ongoing ground-based detector network including LIGO-Hanford, LIGO-Livingston \cite{TheLIGOScientific:2014jea}, Virgo \cite{TheVirgo:2014hva}, and KAGRA \cite{Aso:2013eba}. We need at least five pairs of detectors to algebraically separate the five spectra \cite{Seto:2008sr}. Using the four detectors listed above, we can make $_4C_2 = 6$ pairs in total and can separate the five spectra.  But their estimation errors would be strongly correlated. By adding  LIGO-India \cite{LIGOIndia} to the network,  we will have   $_5C_2 = 10$ pairs of detectors, and the noise correlation  would be considerably reduced. Also, it might be interesting to examine third generation detectors \cite{1865849}.

Another application would be a case study for the space detectors, such as the LISA-Taiji network \cite{Audley:2017drz, Hu:2017mde, Wang:2021mou, Wang:2021uih,Seto:2020zxw, Orlando:2020oko,Omiya:2020fvw}. However, because of the existing geometrical symmetry,  we will  have only three independent data combinations  and cannot  separate the five components completely \cite{Omiya:2020fvw}. If Tian-Qin \cite{Luo:2015ght} is additionally available, we  can  solve the degeneracy in principle.   But, its optimal  frequency  is higher than LISA and Taiji, and the overall performance of the correlation analysis would be limited \cite{Seto:2020mfd}. 

Throughout this paper, we applied the low frequency approximation for responses of individual detectors. This is an efficient approximation for most observational situations, but we have the degeneracy between the two scalar modes. In some cases, we need to carefully deal with the finiteness of the arm length (see e.g. \cite{Seto:2006dz, 1865849}).  It might be interesting to study the possibility of resolving the degeneracy.

\begin{acknowledgments}
We  are sincerely grateful to the referee for carefully reading the draft and pointing out many errors in our expressions. This work is supported by JSPS Kakenhi Grant-in-Aid for Scientific Research
 (Nos. 17H06358 and 19K03870).
\end{acknowledgments}

\appendix
\section{Parity even Overlap reduction functions with orthogonal tensors}\label{sec:AppA}

In this appendix, we derive the ORFs for the even parity spectra ($I_T,I_V,$ and $I_S$) with the method that we applied for the odd parity ones (see Sec. \ref{sec:3B}). The procedure for the irreducible decomposition is almost the same.  But,  for parity even ones, we have the symmetry  \eqref{eq:29} instead of Eq. \eqref{eq:30}. After some calculations, we find that the following five tensors form the orthonormal basis for the decomposition:
\begin{align}
	&H_{ijkl} =  \frac{1}{3}\delta_{ij}\delta_{kl}~,\\
	&K_{ijkl} = \frac{1}{2\sqrt{5}}\left(\delta_{ik}\delta_{jl} + \delta_{il} \delta_{jk}\right) - \frac{1}{\sqrt{5}} H_{ijkl}~,\\
	&H^{0}_{ijkl} = \frac{1}{2}\left(\delta_{ij} M^0_{kl} + M^0_{ij} \delta_{kl}\right)~,\\
	&\begin{aligned}
	K^{0}_{ijkl} =& \frac{3}{2\sqrt{14}}\left(\delta_{ik}M^{0}_{jl} + \delta_{il}M^{0}_{jk}+ \delta_{jk}M^{0}_{il} + \delta_{jl}M^{0}_{ik}\right) \\
	&- \frac{4}{\sqrt{14}}H^{0}_{ijkl}~,
	\end{aligned}\\
	&F^{0}_{ijkl} = \frac{1}{2}\sqrt{\frac{35}{2}}M^{0}_{ij}M^{0}_{kl} - \frac{\sqrt{5}}{3} K^0_{ijkl} -  \frac{1}{3}\sqrt{\frac{7}{2}} K_{ijkl}~.
\end{align}
Note that  $F^0$ satisfies the traceless property
\begin{align}
	F^0_{iijk} = F^0_{ijik} = \dots = F^0_{jkii} = 0~.
\end{align}

Using these basis tensors, the even parity functions (Eq. \eqref{eq:20} - \eqref{eq:22}) are expanded as
\begin{align}
	\Gamma^{I_P}_{ijkl} = &\rho^P_{F^{0}} F^{0}_{ijkl} + \rho^P_{H^{0}} H^{0}_{ijkl} \nonumber\\
	&+ \rho^P_{K^{0}}  K^0_{ijkl} + \rho^P_{H} H_{ijkl} + \rho^P_{K} K_{ijkl} ~.
\end{align}
The orthonormality of the basis tensors allows us to obtain the coefficients as
\begin{align}
	\rho^P_{F^{0}} &= F^{0}_{ijkl}\Gamma^{I_P}_{ijkl}~,\\
	\rho^P_{H^{0}} &= H^{0}_{ijkl}\Gamma^{I_P}_{ijkl}~,\\
	\rho^P_{K^{0}} &= K^{0}_{ijkl}\Gamma^{I_P}_{ijkl}~,\\
	\rho^P_{H} &= H_{ijkl}\Gamma^{I_P}_{ijkl}~,\\
	\rho^P_{K} &= K_{ijkl}\Gamma^{I_P}_{ijkl}~.
\end{align}
After some elementary integral, we obtain
\begin{widetext}
\begin{align}
	(\rho^T_{F^{0}},\rho^T_{H^{0}},\rho^T_{K^{0}},\rho^T_{H},\rho^T_{K}) &= 2\sqrt{5}(\frac{1}{\sqrt{14}} j_4,0, \sqrt{\frac{10}{7}}j_2,0,j_0)~,\\
	(\rho^V_{F^{0}},\rho^V_{H^{0}},\rho^V_{K^{0}},\rho^V_{H},\rho^V_{K}) &= 2\sqrt{5}(-2\sqrt{\frac{2}{7}} j_4,0,-\sqrt{\frac{5}{14}}j_2,0,j_0)~,\\
	(\rho^S_{F^{0}},\rho^S_{H^{0}},\rho^S_{K^{0}},\rho^S_{H},\rho^S_{K}) &= 2\sqrt{5}(3\sqrt{\frac{2}{7}} j_4,\frac{\sqrt{5}}{2} j_2, - \sqrt{\frac{10}{7}}j_2,5\frac{\sqrt{5}}{\add{4}}j_0,j_0)~.
\end{align}
\end{widetext}
By contracting $\Gamma^{I_P}_{ijkl}$ with the detector tensors, we have the explicit form of the ORF as
\begin{widetext}
\begin{align}\label{eq:B16}
	\gamma^{I_T}_{AB} &= 2\sqrt{5}\left(\frac{1}{\sqrt{14}} D^{F}_{AB} j_4(y) \add{+} \sqrt{\frac{10}{7}} D^{K^0}_{AB} j_2(y) +  D^{K}_{AB} j_0(y)\right)~,\\
	\label{eq:B17}
	\gamma^{I_V}_{AB} &= 2\sqrt{5}\left(-2 \sqrt{\frac{2}{7}}D^{F}_{AB} j_4(y) - \sqrt{\frac{5}{14}} D^{K^0}_{AB} j_2(y) +  D^{K}_{AB} j_0(y)\right)~,\\
	\label{eq:B18}
	\gamma^{I_S}_{AB} &= 2\sqrt{5}\left(3\sqrt{\frac{2}{7}} D^{F}_{AB} j_4(y) - \sqrt{\frac{10}{7}} D^{K^0}_{AB} j_2(y) +  D^{K}_{AB} j_0(y)\right)~.
\end{align}
\end{widetext}
Here, we defined
\begin{align}\label{eq:B19}
	D^{F}_{AB} &\equiv D_{A,ij}D_{B,kl} F^0_{ijkl}~,\\
	\label{eq:B20}
	D^{K^0}_{AB} &\equiv D_{A,ij}D_{B,kl} K^0_{ijkl}~,\\
	\label{eq:B21}
	D^{K}_{AB} &\equiv D_{A,ij}D_{B,kl} K_{ijkl}~.	
\end{align}
The coefficients $\rho^S_2$ and $\rho^S_4$ do not contribute to the ORFs, because the contraction with $H_{ijkl}$ and $H^{0}_{ijkl}$ is identically zero
\begin{align}
	D_{A,ij}D_{B,kl} H_{ijkl} = 0~,\\
	D_{A,ij}D_{B,kl} H^0_{ijkl} = 0~,
\end{align}
which obey from the traceless property of the detector tensor.

\section{Explicit formulae for ground-based detector network}\label{sec:AppC}

In this section, we give the explicit formulae of the five ORFs $\gamma^{I_P}$ and $\gamma^{W_P}$ for a network composed by two ground-based detectors.
A ground-based detector is virtually tangential to the Earth\rq{}s surface that can be regarded as a sphere. Therefore, the relative geometry of two arbitrary  detectors is   characterized by the three angles $(\beta,\delta,\Delta)$ (see Fig. 1 and Eq. (21) of \cite{Seto:2008sr} for their definitions). After some algebra, for Eqs. (\ref{eq:51})-(\ref{eq:52}) and (\ref{eq:B19})-(\ref{eq:B21}), we find
\begin{align}
	&D^{\tilde{F}}_{AB} = -\frac{7 + 3 \cos\beta}{8\sqrt{10}}\sin\left(\frac{\beta}{2}\right)\sin4\Delta~, \\
	&D^{\tilde{K}}_{AB} = \frac{1}{\sqrt{10}}\sin^3\left(\frac{\beta}{2}\right)\sin4\Delta~.\\
&\begin{aligned}
	D^{F}_{AB} = &\frac{3}{16\sqrt{70}}\cos^4\left(\frac{\beta}{2}\right) \cos 4\delta \\
	&- \frac{169 + 108\cos\beta + 3 \cos 2\beta}{128\sqrt{70}}\cos4\Delta~,
\end{aligned}\\
	&D^{K^0}_{AB} = \frac{1}{2\sqrt{14}} \cos^4\left(\frac{\beta}{2}\right) \cos 4\delta + \frac{ 5 + \cos\beta }{4 \sqrt{14}}\sin^2\left(\frac{\beta}{2}\right)\cos4\Delta~,\\
	&D^{K}_{AB} =\frac{1}{2\sqrt{5}}\cos^4\left(\frac{\beta}{2}\right) \cos 4\delta - \frac{1}{2\sqrt{5}}\sin^4\left(\frac{\beta}{2}\right)\cos4\Delta~.
\end{align}
Substituting these coefficients to Eqs. \eqref{eq:49}, \eqref{eq:50}, \eqref{eq:B16}, \eqref{eq:B17}, and \eqref{eq:B18}, we obtain
\begin{align}\label{eq:B6}
	\gamma^{W_P}_{} &= \Xi^{P}(y,\beta) \sin4 \Delta~, & (P &= T,V)~,\\
	\gamma^{I_P}_{} &= \Theta_\Delta^{P}(y,\beta) \cos4 \Delta + \Theta_\delta^{P}(y,\beta)\cos4\delta~, & (P &= T,V,S)~.
\end{align}
Here, the coefficients $\Xi^P, \Theta_\Delta^P,$ and $\Theta_\delta^P$ are given by
\begin{widetext}
\begin{align}
		\Xi^T(y,\beta) &= \sin\left(\frac{\beta}{2}\right)\left((1-\cos\beta)j_1(y) - \frac{7 + 3 \cos \beta}{8}j_3(y)\right)~,\\
	\Xi^V(y,\beta) &= \frac{1}{2}\sin\left(\frac{\beta}{2}\right)\left((1-\cos\beta)j_1(y) + \frac{7 + 3 \cos \beta}{2}j_3(y)\right)~,
\end{align}
\begin{align}
	\Theta^T_\Delta(y,\beta) &= - \sin^4 \left(\frac{\beta}{2}\right) j_0(y) - \frac{5}{56}(-9 + 8\cos\beta + \cos2\beta) j_2(y) - \frac{1}{896}(169 + 108 \cos\beta + 3 \cos2\beta) j_4(y)~,\\
	\Theta^V_\Delta(y,\beta) &= - \sin^4 \left(\frac{\beta}{2}\right) j_0(y) + \frac{5}{112}(-9 + 8 \cos\beta  + \cos2\beta) j_2(y) + \frac{1}{224}(169 + 108 \cos\beta + 3 \cos2\beta) j_4(y)~,
\\
	\Theta^S_\Delta(y,\beta) &= - \sin^4 \left(\frac{\beta}{2}\right) j_0(y) + \frac{5}{56}(-9 + 8\cos\beta + \cos 2\beta) j_2(y) - \frac{3}{448}(169 + 108 \cos\beta + 3 \cos2\beta) j_4(y)~,
\end{align}
\begin{align}
	\Theta^T_{\delta}(y,\beta) &= \cos^4\left(\frac{\beta}{2}\right) \left(j_0(y) + \frac{5}{14}j_2(y) + \frac{3}{112} j_4 (y) \right)~,\\
	\Theta^V_{\delta}(y,\beta) &=\cos^4\left(\frac{\beta}{2}\right) \left(j_0(y) - \frac{5}{14}j_2(y) - \frac{3}{28} j_4 (y) \right)~,\\
	\Theta^S_{\delta}(y,\beta) &=\cos^4\left(\frac{\beta}{2}\right) \left(j_0(y) - \frac{5}{7}j_2(y) + \frac{9}{56} j_4 (y) \right)~.
\end{align}
\end{widetext}
These expressions (except for the newly derived $\gamma^{W_V}$) are essentially the same as those in the literature \cite{Flanagan:1993ix, Seto:2008sr, Nishizawa:2009bf}.

\bibliography{ref}

\providecommand{\noopsort}[1]{}\providecommand{\singleletter}[1]{#1}%
\begin{thebibliography}{48}
\expandafter\ifx\csname natexlab\endcsname\relax\def\natexlab#1{#1}\fi
\expandafter\ifx\csname bibnamefont\endcsname\relax
  \def\bibnamefont#1{#1}\fi
\expandafter\ifx\csname bibfnamefont\endcsname\relax
  \def\bibfnamefont#1{#1}\fi
\expandafter\ifx\csname citenamefont\endcsname\relax
  \def\citenamefont#1{#1}\fi
\expandafter\ifx\csname url\endcsname\relax
  \def\url#1{\texttt{#1}}\fi
\expandafter\ifx\csname urlprefix\endcsname\relax\def\urlprefix{URL }\fi
\providecommand{\bibinfo}[2]{#2}
\providecommand{\eprint}[2][]{\url{#2}}

\bibitem[{\citenamefont{Starobinsky}(1979)}]{Starobinsky:1979ty}
\bibinfo{author}{\bibfnamefont{A.~A.} \bibnamefont{Starobinsky}},
  \bibinfo{journal}{JETP Lett.} \textbf{\bibinfo{volume}{30}},
  \bibinfo{pages}{682} (\bibinfo{year}{1979}).

\bibitem[{\citenamefont{Easther et~al.}(2007)\citenamefont{Easther, Giblin, and
  Lim}}]{PhysRevLett.99.221301}
\bibinfo{author}{\bibfnamefont{R.}~\bibnamefont{Easther}},
  \bibinfo{author}{\bibfnamefont{J.~T.} \bibnamefont{Giblin}},
  \bibnamefont{and} \bibinfo{author}{\bibfnamefont{E.~A.} \bibnamefont{Lim}},
  \bibinfo{journal}{Phys. Rev. Lett.} \textbf{\bibinfo{volume}{99}},
  \bibinfo{pages}{221301} (\bibinfo{year}{2007}).

\bibitem[{\citenamefont{Kamionkowski et~al.}(1994)\citenamefont{Kamionkowski,
  Kosowsky, and Turner}}]{Kamionkowski:1993fg}
\bibinfo{author}{\bibfnamefont{M.}~\bibnamefont{Kamionkowski}},
  \bibinfo{author}{\bibfnamefont{A.}~\bibnamefont{Kosowsky}}, \bibnamefont{and}
  \bibinfo{author}{\bibfnamefont{M.~S.} \bibnamefont{Turner}},
  \bibinfo{journal}{Phys. Rev. D} \textbf{\bibinfo{volume}{49}},
  \bibinfo{pages}{2837} (\bibinfo{year}{1994}), \eprint{astro-ph/9310044}.

\bibitem[{\citenamefont{Caprini et~al.}(2008)\citenamefont{Caprini, Durrer, and
  Servant}}]{Caprini:2007xq}
\bibinfo{author}{\bibfnamefont{C.}~\bibnamefont{Caprini}},
  \bibinfo{author}{\bibfnamefont{R.}~\bibnamefont{Durrer}}, \bibnamefont{and}
  \bibinfo{author}{\bibfnamefont{G.}~\bibnamefont{Servant}},
  \bibinfo{journal}{Phys. Rev. D} \textbf{\bibinfo{volume}{77}},
  \bibinfo{pages}{124015} (\bibinfo{year}{2008}), \eprint{0711.2593}.

\bibitem[{\citenamefont{Maggiore}(2000)}]{Maggiore:1999vm}
\bibinfo{author}{\bibfnamefont{M.}~\bibnamefont{Maggiore}},
  \bibinfo{journal}{Phys. Rept.} \textbf{\bibinfo{volume}{331}},
  \bibinfo{pages}{283} (\bibinfo{year}{2000}), \eprint{gr-qc/9909001}.

\bibitem[{\citenamefont{Romano and Cornish}(2017)}]{Romano:2016dpx}
\bibinfo{author}{\bibfnamefont{J.~D.} \bibnamefont{Romano}} \bibnamefont{and}
  \bibinfo{author}{\bibfnamefont{N.~J.} \bibnamefont{Cornish}},
  \bibinfo{journal}{Living Rev. Rel.} \textbf{\bibinfo{volume}{20}},
  \bibinfo{pages}{2} (\bibinfo{year}{2017}), \eprint{1608.06889}.

\bibitem[{\citenamefont{Christensen}(2019)}]{Christensen:2018iqi}
\bibinfo{author}{\bibfnamefont{N.}~\bibnamefont{Christensen}},
  \bibinfo{journal}{Rept. Prog. Phys.} \textbf{\bibinfo{volume}{82}},
  \bibinfo{pages}{016903} (\bibinfo{year}{2019}), \eprint{1811.08797}.

\bibitem[{\citenamefont{Kuroyanagi et~al.}(2018)\citenamefont{Kuroyanagi,
  Chiba, and Takahashi}}]{Kuroyanagi:2018csn}
\bibinfo{author}{\bibfnamefont{S.}~\bibnamefont{Kuroyanagi}},
  \bibinfo{author}{\bibfnamefont{T.}~\bibnamefont{Chiba}}, \bibnamefont{and}
  \bibinfo{author}{\bibfnamefont{T.}~\bibnamefont{Takahashi}},
  \bibinfo{journal}{JCAP} \textbf{\bibinfo{volume}{11}}, \bibinfo{pages}{038}
  (\bibinfo{year}{2018}), \eprint{1807.00786}.

\bibitem[{\citenamefont{Will}(1993)}]{Will:1993ns}
\bibinfo{author}{\bibfnamefont{C.}~\bibnamefont{Will}},
  \emph{\bibinfo{title}{{Theory and experiment in gravitational physics}}}
  (\bibinfo{year}{1993}), ISBN \bibinfo{isbn}{978-0-521-43973-2}.

\bibitem[{\citenamefont{Nishizawa et~al.}(2009)\citenamefont{Nishizawa, Taruya,
  Hayama, Kawamura, and Sakagami}}]{Nishizawa:2009bf}
\bibinfo{author}{\bibfnamefont{A.}~\bibnamefont{Nishizawa}},
  \bibinfo{author}{\bibfnamefont{A.}~\bibnamefont{Taruya}},
  \bibinfo{author}{\bibfnamefont{K.}~\bibnamefont{Hayama}},
  \bibinfo{author}{\bibfnamefont{S.}~\bibnamefont{Kawamura}}, \bibnamefont{and}
  \bibinfo{author}{\bibfnamefont{M.-a.} \bibnamefont{Sakagami}},
  \bibinfo{journal}{Phys. Rev. D} \textbf{\bibinfo{volume}{79}},
  \bibinfo{pages}{082002} (\bibinfo{year}{2009}), \eprint{0903.0528}.

\bibitem[{\citenamefont{Nishizawa et~al.}(2010)\citenamefont{Nishizawa, Taruya,
  and Kawamura}}]{Nishizawa:2009jh}
\bibinfo{author}{\bibfnamefont{A.}~\bibnamefont{Nishizawa}},
  \bibinfo{author}{\bibfnamefont{A.}~\bibnamefont{Taruya}}, \bibnamefont{and}
  \bibinfo{author}{\bibfnamefont{S.}~\bibnamefont{Kawamura}},
  \bibinfo{journal}{Phys. Rev. D} \textbf{\bibinfo{volume}{81}},
  \bibinfo{pages}{104043} (\bibinfo{year}{2010}), \eprint{0911.0525}.

\bibitem[{\citenamefont{Cornish et~al.}(2018)\citenamefont{Cornish, O'Beirne,
  Taylor, and Yunes}}]{Cornish:2017oic}
\bibinfo{author}{\bibfnamefont{N.~J.} \bibnamefont{Cornish}},
  \bibinfo{author}{\bibfnamefont{L.}~\bibnamefont{O'Beirne}},
  \bibinfo{author}{\bibfnamefont{S.~R.} \bibnamefont{Taylor}},
  \bibnamefont{and} \bibinfo{author}{\bibfnamefont{N.}~\bibnamefont{Yunes}},
  \bibinfo{journal}{Phys. Rev. Lett.} \textbf{\bibinfo{volume}{120}},
  \bibinfo{pages}{181101} (\bibinfo{year}{2018}), \eprint{1712.07132}.

\bibitem[{\citenamefont{Abbott et~al.}(2019)}]{LIGOScientific:2019vic}
\bibinfo{author}{\bibfnamefont{B.}~\bibnamefont{Abbott}} \bibnamefont{et~al.}
  (\bibinfo{collaboration}{LIGO Scientific, Virgo}), \bibinfo{journal}{Phys.
  Rev. D} \textbf{\bibinfo{volume}{100}}, \bibinfo{pages}{061101}
  (\bibinfo{year}{2019}), \eprint{1903.02886}.

\bibitem[{\citenamefont{Lue et~al.}(1999)\citenamefont{Lue, Wang, and
  Kamionkowski}}]{Lue:1998mq}
\bibinfo{author}{\bibfnamefont{A.}~\bibnamefont{Lue}},
  \bibinfo{author}{\bibfnamefont{L.-M.} \bibnamefont{Wang}}, \bibnamefont{and}
  \bibinfo{author}{\bibfnamefont{M.}~\bibnamefont{Kamionkowski}},
  \bibinfo{journal}{Phys. Rev. Lett.} \textbf{\bibinfo{volume}{83}},
  \bibinfo{pages}{1506} (\bibinfo{year}{1999}), \eprint{astro-ph/9812088}.

\bibitem[{\citenamefont{Seto}(2006)}]{Seto:2006hf}
\bibinfo{author}{\bibfnamefont{N.}~\bibnamefont{Seto}}, \bibinfo{journal}{Phys.
  Rev. Lett.} \textbf{\bibinfo{volume}{97}}, \bibinfo{pages}{151101}
  (\bibinfo{year}{2006}), \eprint{astro-ph/0609504}.

\bibitem[{\citenamefont{Kato and Soda}(2016)}]{Kato:2015bye}
\bibinfo{author}{\bibfnamefont{R.}~\bibnamefont{Kato}} \bibnamefont{and}
  \bibinfo{author}{\bibfnamefont{J.}~\bibnamefont{Soda}},
  \bibinfo{journal}{Phys. Rev. D} \textbf{\bibinfo{volume}{93}},
  \bibinfo{pages}{062003} (\bibinfo{year}{2016}), \eprint{1512.09139}.

\bibitem[{\citenamefont{Smith and Caldwell}(2017)}]{Smith:2016jqs}
\bibinfo{author}{\bibfnamefont{T.~L.} \bibnamefont{Smith}} \bibnamefont{and}
  \bibinfo{author}{\bibfnamefont{R.}~\bibnamefont{Caldwell}},
  \bibinfo{journal}{Phys. Rev. D} \textbf{\bibinfo{volume}{95}},
  \bibinfo{pages}{044036} (\bibinfo{year}{2017}), \eprint{1609.05901}.

\bibitem[{\citenamefont{Domcke et~al.}(2020)\citenamefont{Domcke,
  Garcia-Bellido, Peloso, Pieroni, Ricciardone, Sorbo, and
  Tasinato}}]{Domcke:2019zls}
\bibinfo{author}{\bibfnamefont{V.}~\bibnamefont{Domcke}},
  \bibinfo{author}{\bibfnamefont{J.}~\bibnamefont{Garcia-Bellido}},
  \bibinfo{author}{\bibfnamefont{M.}~\bibnamefont{Peloso}},
  \bibinfo{author}{\bibfnamefont{M.}~\bibnamefont{Pieroni}},
  \bibinfo{author}{\bibfnamefont{A.}~\bibnamefont{Ricciardone}},
  \bibinfo{author}{\bibfnamefont{L.}~\bibnamefont{Sorbo}}, \bibnamefont{and}
  \bibinfo{author}{\bibfnamefont{G.}~\bibnamefont{Tasinato}},
  \bibinfo{journal}{JCAP} \textbf{\bibinfo{volume}{05}}, \bibinfo{pages}{028}
  (\bibinfo{year}{2020}), \eprint{1910.08052}.

\bibitem[{\citenamefont{Belgacem and Kamionkowski}(2020)}]{Belgacem:2020nda}
\bibinfo{author}{\bibfnamefont{E.}~\bibnamefont{Belgacem}} \bibnamefont{and}
  \bibinfo{author}{\bibfnamefont{M.}~\bibnamefont{Kamionkowski}},
  \bibinfo{journal}{Phys. Rev. D} \textbf{\bibinfo{volume}{102}},
  \bibinfo{pages}{023004} (\bibinfo{year}{2020}), \eprint{2004.05480}.

\bibitem[{\citenamefont{Alexander et~al.}(2006)\citenamefont{Alexander, Peskin,
  and Sheikh-Jabbari}}]{Alexander:2004us}
\bibinfo{author}{\bibfnamefont{S.~H.-S.} \bibnamefont{Alexander}},
  \bibinfo{author}{\bibfnamefont{M.~E.} \bibnamefont{Peskin}},
  \bibnamefont{and} \bibinfo{author}{\bibfnamefont{M.~M.}
  \bibnamefont{Sheikh-Jabbari}}, \bibinfo{journal}{Phys. Rev. Lett.}
  \textbf{\bibinfo{volume}{96}}, \bibinfo{pages}{081301}
  (\bibinfo{year}{2006}), \eprint{hep-th/0403069}.

\bibitem[{\citenamefont{Satoh et~al.}(2008)\citenamefont{Satoh, Kanno, and
  Soda}}]{Satoh:2007gn}
\bibinfo{author}{\bibfnamefont{M.}~\bibnamefont{Satoh}},
  \bibinfo{author}{\bibfnamefont{S.}~\bibnamefont{Kanno}}, \bibnamefont{and}
  \bibinfo{author}{\bibfnamefont{J.}~\bibnamefont{Soda}},
  \bibinfo{journal}{Phys. Rev. D} \textbf{\bibinfo{volume}{77}},
  \bibinfo{pages}{023526} (\bibinfo{year}{2008}), \eprint{0706.3585}.

\bibitem[{\citenamefont{Adshead and Wyman}(2012)}]{Adshead:2012kp}
\bibinfo{author}{\bibfnamefont{P.}~\bibnamefont{Adshead}} \bibnamefont{and}
  \bibinfo{author}{\bibfnamefont{M.}~\bibnamefont{Wyman}},
  \bibinfo{journal}{Phys. Rev. Lett.} \textbf{\bibinfo{volume}{108}},
  \bibinfo{pages}{261302} (\bibinfo{year}{2012}), \eprint{1202.2366}.

\bibitem[{\citenamefont{Kahniashvili et~al.}(2005)\citenamefont{Kahniashvili,
  Gogoberidze, and Ratra}}]{Kahniashvili:2005qi}
\bibinfo{author}{\bibfnamefont{T.}~\bibnamefont{Kahniashvili}},
  \bibinfo{author}{\bibfnamefont{G.}~\bibnamefont{Gogoberidze}},
  \bibnamefont{and} \bibinfo{author}{\bibfnamefont{B.}~\bibnamefont{Ratra}},
  \bibinfo{journal}{Phys. Rev. Lett.} \textbf{\bibinfo{volume}{95}},
  \bibinfo{pages}{151301} (\bibinfo{year}{2005}), \eprint{astro-ph/0505628}.

\bibitem[{\citenamefont{Ellis et~al.}(2020)\citenamefont{Ellis, Fairbairn,
  Lewicki, Vaskonen, and Wickens}}]{Ellis:2020uid}
\bibinfo{author}{\bibfnamefont{J.}~\bibnamefont{Ellis}},
  \bibinfo{author}{\bibfnamefont{M.}~\bibnamefont{Fairbairn}},
  \bibinfo{author}{\bibfnamefont{M.}~\bibnamefont{Lewicki}},
  \bibinfo{author}{\bibfnamefont{V.}~\bibnamefont{Vaskonen}}, \bibnamefont{and}
  \bibinfo{author}{\bibfnamefont{A.}~\bibnamefont{Wickens}},
  \bibinfo{journal}{JCAP} \textbf{\bibinfo{volume}{10}}, \bibinfo{pages}{032}
  (\bibinfo{year}{2020}), \eprint{2005.05278}.

\bibitem[{\citenamefont{Christensen}(1992)}]{Christensen:1992wi}
\bibinfo{author}{\bibfnamefont{N.}~\bibnamefont{Christensen}},
  \bibinfo{journal}{Phys. Rev. D} \textbf{\bibinfo{volume}{46}},
  \bibinfo{pages}{5250} (\bibinfo{year}{1992}).

\bibitem[{\citenamefont{Flanagan}(1993)}]{Flanagan:1993ix}
\bibinfo{author}{\bibfnamefont{E.~E.} \bibnamefont{Flanagan}},
  \bibinfo{journal}{Phys. Rev. D} \textbf{\bibinfo{volume}{48}},
  \bibinfo{pages}{2389} (\bibinfo{year}{1993}), \eprint{astro-ph/9305029}.

\bibitem[{\citenamefont{Allen and Romano}(1999)}]{Allen:1997ad}
\bibinfo{author}{\bibfnamefont{B.}~\bibnamefont{Allen}} \bibnamefont{and}
  \bibinfo{author}{\bibfnamefont{J.~D.} \bibnamefont{Romano}},
  \bibinfo{journal}{Phys. Rev. D} \textbf{\bibinfo{volume}{59}},
  \bibinfo{pages}{102001} (\bibinfo{year}{1999}), \eprint{gr-qc/9710117}.

\bibitem[{\citenamefont{Seto and Taruya}(2008)}]{Seto:2008sr}
\bibinfo{author}{\bibfnamefont{N.}~\bibnamefont{Seto}} \bibnamefont{and}
  \bibinfo{author}{\bibfnamefont{A.}~\bibnamefont{Taruya}},
  \bibinfo{journal}{Phys. Rev. D} \textbf{\bibinfo{volume}{77}},
  \bibinfo{pages}{103001} (\bibinfo{year}{2008}), \eprint{0801.4185}.

\bibitem[{\citenamefont{Abbott et~al.}(2017)}]{TheLIGOScientific:2017qsa}
\bibinfo{author}{\bibfnamefont{B.}~\bibnamefont{Abbott}} \bibnamefont{et~al.}
  (\bibinfo{collaboration}{LIGO Scientific, Virgo}), \bibinfo{journal}{Phys.
  Rev. Lett.} \textbf{\bibinfo{volume}{119}}, \bibinfo{pages}{161101}
  (\bibinfo{year}{2017}), \eprint{1710.05832}.

\bibitem[{\citenamefont{Omiya and Seto}(2020)}]{Omiya:2020fvw}
\bibinfo{author}{\bibfnamefont{H.}~\bibnamefont{Omiya}} \bibnamefont{and}
  \bibinfo{author}{\bibfnamefont{N.}~\bibnamefont{Seto}},
  \bibinfo{journal}{Phys. Rev. D} \textbf{\bibinfo{volume}{102}},
  \bibinfo{pages}{084053} (\bibinfo{year}{2020}), \eprint{2010.00771}.

\bibitem[{\citenamefont{{Rybicki} and {Lightman}}(1979)}]{1979rpa..book.....R}
\bibinfo{author}{\bibfnamefont{G.~B.} \bibnamefont{{Rybicki}}}
  \bibnamefont{and} \bibinfo{author}{\bibfnamefont{A.~P.}
  \bibnamefont{{Lightman}}}, \emph{\bibinfo{title}{{Radiative processes in
  astrophysics}}} (\bibinfo{year}{1979}).

\bibitem[{\citenamefont{Seto}(2007)}]{Seto:2006dz}
\bibinfo{author}{\bibfnamefont{N.}~\bibnamefont{Seto}}, \bibinfo{journal}{Phys.
  Rev. D} \textbf{\bibinfo{volume}{75}}, \bibinfo{pages}{061302}
  (\bibinfo{year}{2007}), \eprint{astro-ph/0609633}.

\bibitem[{\citenamefont{Seto and Taruya}(2007)}]{Seto:2007tn}
\bibinfo{author}{\bibfnamefont{N.}~\bibnamefont{Seto}} \bibnamefont{and}
  \bibinfo{author}{\bibfnamefont{A.}~\bibnamefont{Taruya}},
  \bibinfo{journal}{Phys. Rev. Lett.} \textbf{\bibinfo{volume}{99}},
  \bibinfo{pages}{121101} (\bibinfo{year}{2007}), \eprint{0707.0535}.

\bibitem[{\citenamefont{Forward}(1978)}]{Forward:1978zm}
\bibinfo{author}{\bibfnamefont{R.~L.} \bibnamefont{Forward}},
  \bibinfo{journal}{Phys. Rev. D} \textbf{\bibinfo{volume}{17}},
  \bibinfo{pages}{379} (\bibinfo{year}{1978}).

\bibitem[{\citenamefont{Hamermesh}(1989)}]{hamermesh1989group}
\bibinfo{author}{\bibfnamefont{M.}~\bibnamefont{Hamermesh}},
  \emph{\bibinfo{title}{Group Theory and Its Application to Physical
  Problems}}, Addison Wesley Series in Physics (\bibinfo{publisher}{Dover
  Publications}, \bibinfo{year}{1989}), ISBN \bibinfo{isbn}{9780486661810},
  \urlprefix\url{https://books.google.co.jp/books?id=c0o9\_wlCzgcC}.

\bibitem[{\citenamefont{Aasi et~al.}(2015)}]{TheLIGOScientific:2014jea}
\bibinfo{author}{\bibfnamefont{J.}~\bibnamefont{Aasi}} \bibnamefont{et~al.}
  (\bibinfo{collaboration}{LIGO Scientific}), \bibinfo{journal}{Class. Quant.
  Grav.} \textbf{\bibinfo{volume}{32}}, \bibinfo{pages}{074001}
  (\bibinfo{year}{2015}), \eprint{1411.4547}.

\bibitem[{\citenamefont{Acernese et~al.}(2015)}]{TheVirgo:2014hva}
\bibinfo{author}{\bibfnamefont{F.}~\bibnamefont{Acernese}} \bibnamefont{et~al.}
  (\bibinfo{collaboration}{VIRGO}), \bibinfo{journal}{Class. Quant. Grav.}
  \textbf{\bibinfo{volume}{32}}, \bibinfo{pages}{024001}
  (\bibinfo{year}{2015}), \eprint{1408.3978}.

\bibitem[{\citenamefont{Aso et~al.}(2013)\citenamefont{Aso, Michimura, Somiya,
  Ando, Miyakawa, Sekiguchi, Tatsumi, and Yamamoto}}]{Aso:2013eba}
\bibinfo{author}{\bibfnamefont{Y.}~\bibnamefont{Aso}},
  \bibinfo{author}{\bibfnamefont{Y.}~\bibnamefont{Michimura}},
  \bibinfo{author}{\bibfnamefont{K.}~\bibnamefont{Somiya}},
  \bibinfo{author}{\bibfnamefont{M.}~\bibnamefont{Ando}},
  \bibinfo{author}{\bibfnamefont{O.}~\bibnamefont{Miyakawa}},
  \bibinfo{author}{\bibfnamefont{T.}~\bibnamefont{Sekiguchi}},
  \bibinfo{author}{\bibfnamefont{D.}~\bibnamefont{Tatsumi}}, \bibnamefont{and}
  \bibinfo{author}{\bibfnamefont{H.}~\bibnamefont{Yamamoto}}
  (\bibinfo{collaboration}{KAGRA}), \bibinfo{journal}{Phys. Rev. D}
  \textbf{\bibinfo{volume}{88}}, \bibinfo{pages}{043007}
  (\bibinfo{year}{2013}), \eprint{1306.6747}.

\bibitem[{\citenamefont{Bala et~al.}(2011)\citenamefont{Bala, Tarun, CS,
  Sanjeev, Sendhil, and Anand}}]{LIGOIndia}
\bibinfo{author}{\bibfnamefont{I.}~\bibnamefont{Bala}},
  \bibinfo{author}{\bibfnamefont{S.}~\bibnamefont{Tarun}},
  \bibinfo{author}{\bibfnamefont{U.}~\bibnamefont{CS}},
  \bibinfo{author}{\bibfnamefont{D.}~\bibnamefont{Sanjeev}},
  \bibinfo{author}{\bibfnamefont{R.}~\bibnamefont{Sendhil}}, \bibnamefont{and}
  \bibinfo{author}{\bibfnamefont{S.}~\bibnamefont{Anand}}
  (\bibinfo{year}{2011}),
  \urlprefix\url{https://dcc.ligo.org/LIGO-M1100296/public}.

\bibitem[{\citenamefont{Amalberti et~al.}(2021)\citenamefont{Amalberti,
  Bartolo, and Ricciardone}}]{1865849}
\bibinfo{author}{\bibfnamefont{L.}~\bibnamefont{Amalberti}},
  \bibinfo{author}{\bibfnamefont{N.}~\bibnamefont{Bartolo}}, \bibnamefont{and}
  \bibinfo{author}{\bibfnamefont{A.}~\bibnamefont{Ricciardone}}
  (\bibinfo{year}{2021}), \eprint{2105.13197}.

\bibitem[{\citenamefont{Amaro-Seoane et~al.}(2017)}]{Audley:2017drz}
\bibinfo{author}{\bibfnamefont{P.}~\bibnamefont{Amaro-Seoane}}
  \bibnamefont{et~al.} (\bibinfo{collaboration}{LISA}) (\bibinfo{year}{2017}),
  \eprint{1702.00786}.

\bibitem[{\citenamefont{Hu and Wu}(2017)}]{Hu:2017mde}
\bibinfo{author}{\bibfnamefont{W.-R.} \bibnamefont{Hu}} \bibnamefont{and}
  \bibinfo{author}{\bibfnamefont{Y.-L.} \bibnamefont{Wu}},
  \bibinfo{journal}{Natl. Sci. Rev.} \textbf{\bibinfo{volume}{4}},
  \bibinfo{pages}{685} (\bibinfo{year}{2017}).

\bibitem[{\citenamefont{Wang and Han}(2021)}]{Wang:2021mou}
\bibinfo{author}{\bibfnamefont{G.}~\bibnamefont{Wang}} \bibnamefont{and}
  \bibinfo{author}{\bibfnamefont{W.-B.} \bibnamefont{Han}},
  \bibinfo{journal}{Phys. Rev. D} \textbf{\bibinfo{volume}{103}},
  \bibinfo{pages}{064021} (\bibinfo{year}{2021}), \eprint{2101.01991}.

\bibitem[{\citenamefont{Wang et~al.}(2021)\citenamefont{Wang, Ni, Han, and
  Xu}}]{Wang:2021uih}
\bibinfo{author}{\bibfnamefont{G.}~\bibnamefont{Wang}},
  \bibinfo{author}{\bibfnamefont{W.-T.} \bibnamefont{Ni}},
  \bibinfo{author}{\bibfnamefont{W.-B.} \bibnamefont{Han}}, \bibnamefont{and}
  \bibinfo{author}{\bibfnamefont{P.}~\bibnamefont{Xu}} (\bibinfo{year}{2021}),
  \eprint{2105.00746}.

\bibitem[{\citenamefont{Seto}(2020{\natexlab{a}})}]{Seto:2020zxw}
\bibinfo{author}{\bibfnamefont{N.}~\bibnamefont{Seto}}, \bibinfo{journal}{Phys.
  Rev. Lett.} \textbf{\bibinfo{volume}{125}}, \bibinfo{pages}{251101}
  (\bibinfo{year}{2020}{\natexlab{a}}), \eprint{2009.02928}.

\bibitem[{\citenamefont{Orlando et~al.}(2021)\citenamefont{Orlando, Pieroni,
  and Ricciardone}}]{Orlando:2020oko}
\bibinfo{author}{\bibfnamefont{G.}~\bibnamefont{Orlando}},
  \bibinfo{author}{\bibfnamefont{M.}~\bibnamefont{Pieroni}}, \bibnamefont{and}
  \bibinfo{author}{\bibfnamefont{A.}~\bibnamefont{Ricciardone}},
  \bibinfo{journal}{JCAP} \textbf{\bibinfo{volume}{03}}, \bibinfo{pages}{069}
  (\bibinfo{year}{2021}), \eprint{2011.07059}.

\bibitem[{\citenamefont{Luo et~al.}(2016)}]{Luo:2015ght}
\bibinfo{author}{\bibfnamefont{J.}~\bibnamefont{Luo}} \bibnamefont{et~al.}
  (\bibinfo{collaboration}{TianQin}), \bibinfo{journal}{Class. Quant. Grav.}
  \textbf{\bibinfo{volume}{33}}, \bibinfo{pages}{035010}
  (\bibinfo{year}{2016}), \eprint{1512.02076}.

\bibitem[{\citenamefont{Seto}(2020{\natexlab{b}})}]{Seto:2020mfd}
\bibinfo{author}{\bibfnamefont{N.}~\bibnamefont{Seto}}, \bibinfo{journal}{Phys.
  Rev. D} \textbf{\bibinfo{volume}{102}}, \bibinfo{pages}{123547}
  (\bibinfo{year}{2020}{\natexlab{b}}), \eprint{2010.06877}.

\end{thebibliography}

\end{document}